\documentclass[reprint,aps,prl,amsmath,amssymb,nofootinbib,superscriptaddress]{revtex4-2}

\usepackage[utf8]{inputenc}
\usepackage[T1]{fontenc}

\usepackage{amsmath}
\usepackage{txfonts}
\usepackage{mathtools}
\usepackage{braket}
\usepackage{tensor}
\usepackage{xcolor}
\usepackage[abbreviations]{siunitx}
\usepackage{graphicx}
\usepackage{booktabs}

\usepackage{float}

\usepackage{hyperref}
\usepackage{cleveref}

\DeclareSIUnit{\fm}{\femto\meter}
\DeclareSIUnit{\MeVc}{\MeV\per\text{\ensuremath{c}}}

\newcommand{\znubb}{\ensuremath{0\nu\beta\beta}}

\allowdisplaybreaks[3]

\begin{document}
\title{Ab initio calculation of the contact operator contribution in the standard mechanism for neutrinoless double beta decay}

\author{R. Wirth}
\email{wirth@frib.msu.edu}
\affiliation{Facility for Rare Isotope Beams, Michigan State University, East Lansing, Michigan 48824-1321, USA}

\author{J. M. Yao}
\email{yaojm8@mail.sysu.edu.cn}
\affiliation{School of Physics and Astronomy, Sun Yat-sen University, Zhuhai 519082, P.R. China}
\affiliation{Facility for Rare Isotope Beams, Michigan State University, East Lansing, Michigan 48824-1321, USA}

\author{H. Hergert}
\email{hergert@frib.msu.edu}
\affiliation{Facility for Rare Isotope Beams, Michigan State University, East Lansing, Michigan 48824-1321, USA}
\affiliation{Department of Physics \& Astronomy, Michigan State University, East Lansing, Michigan 48824-1321, USA}


\begin{abstract}
 Starting from chiral nuclear interactions, we evaluate the contribution of the leading-order contact transition operator to the nuclear matrix element (NME) of neutrinoless double-beta decay, assuming a light Majorana neutrino-exchange mechanism.  The corresponding low-energy constant (LEC) is determined by fitting the transition amplitude of the $nn\to ppe^-e^-$ process to a recently proposed synthetic datum.  We examine the dependence of the amplitude on similarity renormalization group (SRG) scale and chiral expansion order of the nuclear interaction, finding that both dependences can be compensated to a large extent by readjusting the LEC.
 We evaluate the contribution of both the leading-order contact operator and standard long-range operator to the neutrinoless double-beta decays in the light nuclei $\nuclide[6,8]{He}$ and the candidate nucleus $\nuclide[48]{Ca}$.
 Our results provide the first clear demonstration that the contact term enhances the NME by \SI{43(7)}{\percent} in $\nuclide[48]{Ca}$, where the uncertainty is propagated from the synthetic datum.
\end{abstract}

\maketitle

\paragraph{Introduction.} The neutrinoless double-$\beta$ ($\znubb$) decay is a hypothetical weak process that converts two neutrons into two protons, emitting two electrons but no corresponding antineutrinos.
The observation of neutrino oscillations confirmed that neutrinos have nonzero masses, which have boosted interest in experimental searches for $\znubb$ decay.
The observation of this decay would confirm the existence of a Majorana mass term for the neutrinos \cite{Schechter:1982}, shedding light on the mechanism of neutrino mass generation, and providing direct evidence of lepton number violation beyond the standard model---a key ingredient for generating the matter-antimatter asymmetry in the universe.
Hence, there is a vast interest in this process with multiple large-scale experiments planned or underway searching for it.

An important ingredient used to find suitable candidate nuclei for the search, and to interpret an observed lifetime is the nuclear matrix element (NME) $M^{0\nu}$ which contains the part of the decay that is governed by the structure of the parent and daughter nuclei.
There are many calculations of the NME for candidate nuclei in different frameworks~\cite{Menendez:2009,Rodriguez:2010,Barea:2013,Mustonen:2013,Holt:2013,Kwiatkowski:2014,Song:2014,Yao:2015,Hyvarinen:2015,Horoi:2016,Song:2017,Jiao:2017,Yoshinaga:2018,Fang:2018,Rath:2019,Terasaki:2019,Coraggio:2020,Deppisch:2020ztt}, with results differing by up to a factor of 3.
This difference amounts to an order of magnitude uncertainty in the lifetime, the inverse of which is proportional to the square of the NME \cite{em}.

To overcome this, a systematic calculation with quantified uncertainties is required.
Such a calculation can be carried out in an \emph{ab initio} framework using chiral effective field theory (EFT) to derive the nuclear Hamiltonian and $\znubb$ transition operator in a systematically improvable manner.
First milestone calculations have been performed over the previous years, covering several candidate nuclei from \nuclide[48]{Ca} to \nuclide[82]{Se} \cite{Yao2020,Belley2021,Novario:2020dmr}.

Recently, it has been realized that a chiral effective field theory describing $\znubb$ decay based on the mechanism of light Majorana neutrino exchange requires a contact term at leading order that contributes to the decay operator in order to ensure renormalizability \cite{Cirigliano:2018,Cirigliano2019}.
The strength of this contact term has to be determined by matching to a fundamental theory or experimental data.
In the absence of experimental data, only the former is currently possible.

Cirigliano \emph{et al.}~\cite{Cirigliano2020,Cirigliano2021} proposed a way to estimate the size of the contact term by computing the $nn\to ppe^-e^-$ transition amplitude using the generalized forward Compton scattering amplitude.
The underlying model assumes light Majorana neutrino exchange and incorporates input from elastic intermediate states in analogy to the Cottingham formula~\cite{Cottingham1963}.
Since the strength of the contact term is scale and scheme dependent, they provide the value of the full transition amplitude at a given kinematic point.
This amplitude is (in principle) observable, and can be used as a synthetic datum to constrain the contact term in other schemes.

In this work, we compute the $nn\to ppe^-e^-$ transition amplitude using chiral nucleon-nucleon (NN) interactions.
We show that the renormalized transition amplitude is robust with respect to changes in the nuclear interaction, making it a reliable starting point for NME calculations in finite nuclei.
In particular, we investigate the change of the contact contribution when subjecting the NN interaction to a similarity renormalization group (SRG) transformation, as well as its dependence on the expansion order of a chiral interaction.
Finally, we show that the leading-order contact transition operator enhances the NME by \SI{43(7)}{\percent} in the lightest $\znubb$-decay candidate nucleus $\nuclide[48]{Ca}$ compared to the recent ab initio calculations with only the standard long-range transition operator~\cite{Yao2020}.
This finding conveys an important positive message for planning and interpreting future experiments.

\paragraph{The \texorpdfstring{$\znubb$}{0vββ} decay operators.}
The central object of our investigation is the $\znubb$ transition operator in the standard light Majorana neutrino-exchange mechanism.
Since the transition amplitude is computed in the $^1S_0$ channel, we restrict the discussion of the operators to that channel, which simplifies the resulting expressions.
In particular, the only contributing parts of the neutrino potentials are the Fermi (F) and Gamow-Teller (GT) parts.
The tensor part does not contribute in this channel.

The leading-order neutrino potentials in the $^1S_0$ channel are given by
\begin{align}
  V_F(r) &= -\frac{g_V^2}{4\pi r}, \\
  V_{GT}(r) &= -\frac{g_A^2}{4\pi r} \biggl[3 - e^{-m_\pi r} \biggl(1 + \frac{m_\pi r}{2}\biggr)\biggr].
\end{align}
We use the axial coupling constant $g_A=1.27$ and the average pion mass $m_\pi=\SI{138.039}{\MeV}$.
Phenomenological and higher-order corrections can be incorporated in momentum space,
\begin{align}\label{eq:neutrino pheno}
  \tilde V_i(r) &= \frac{1}{2\pi^2} \int_0^\infty \mathrm{d}q \, q^2 h_i(q) j_0(qr),\\
\shortintertext{where}
  h_F(q) &= - \frac{g^2_V(q)}{q^2} \\
  h_{GT}(q) &= -3 \frac{g^2_A(q)}{q^2} - \frac{q^2 g_P^2(q)}{4m_N^2} - \frac{g_A(q) g_P(q)}{m_N} - \frac{g^2_M(q)}{2m_N^2}.
\end{align}
The couplings contain dipole form factors and are given by
\begin{align}
  g_V(q) &= g_V \biggl(1 + \frac{q^2}{\Lambda_V^2}\biggr)^{-2}, &
  g_A(q) &= g_A \biggl(1 + \frac{q^2}{\Lambda_A^2}\biggr)^{-2}, \\
  g_M(q) &= (1+\kappa_1) g_V(q), &
  g_P(q) &= -\frac{2 m_N g_A(q)}{q^2 + m_\pi^2}.
\end{align}
Here, $m_N = \SI{938.919}{\MeV}$ is the average nucleon mass, the symbol $\kappa_1 = 3.7$ is the isovector anomalous magnetic moment of the nucleon.
Following Ref.~\cite{Simkovic:1999}, we choose $\Lambda_V=\SI{850}{\MeVc}$ and $\Lambda_A=\SI{1090}{\MeVc}$ for the vector and axial form factors.
These corrections modify the potentials at short range only.
At longer range, the potentials are identical to the leading-order ones.

The short-range part needed to renormalize the operator is given by a nonlocally regularized contact interaction,
\begin{equation}
  h_S(p, p') = \biggl(\frac{m_N g_A^2}{4f_\pi^2}\biggr)^2 \exp\biggl[-\biggl(\frac{p}{\Lambda}\biggr)^{2n_\text{exp}}\biggr]\exp\biggl[-\biggl(\frac{p'}{\Lambda}\biggr)^{2n_\text{exp}}\biggr],
\end{equation}
with the pion decay constant $f_\pi=\SI{92.2}{\MeV}$.
The contact interaction can be expressed in coordinate space as
\begin{align}
  V_S(r,r') &= \biggl(\frac{m_N g_A^2}{4f_\pi^2}\biggr)^2 f_\Lambda^{n_\text{exp}}(r) f_\Lambda^{n_\text{exp}}(r'),
\shortintertext{where}
  f_\Lambda^{n_\text{exp}}(r) &= \frac{1}{2\pi^2} \int_0^\infty \mathrm{d}q\, q^2 \exp\biggl[-\biggl(\frac{q}{\Lambda}\biggr)^{2n_\text{exp}}\biggr] j_0(qr).
\end{align}
We choose the prefactor similar to Ref.~\cite{Cirigliano2019}, such that the LEC multiplying the contact term becomes dimensionless and of natural size.

The definition of the nuclear part of the $nn \to ppe^-e^-$ transition amplitude compatible with Refs.~\cite{Cirigliano2020,Cirigliano2021} is
\begin{equation}
  \mathcal{A}(p,p') = 4\pi \braket{{}^1S_0(p') | \hat V_F + \hat V_{GT} - 2 g \hat V_S | {}^1S_0(p)}.
\end{equation}
The wavefunctions $\ket{{}^1S_0(p)}$ and $\ket{{}^1S_0(p')}$ are scattering solutions for neutrons and protons in the $^1S_0$ channel at incoming and outgoing momenta $p$ and $p'$, respectively.
Analogously, we define amplitudes $\tilde{\mathcal{A}}(p,p')$ using the neutrino potentials from \cref{eq:neutrino pheno}.
In that case, we denote the LEC multiplying the short-range operator by $\tilde{g}$.

\paragraph{Scattering wavefunctions.}
We compute scattering wavefunctions using the $R$-matrix formalism \cite{Descouvemont2010} with the channel radius set to $a=\SI{15}{\fm}$, well beyond the range of the nuclear potential.
The wavefunctions are normalized such that the asymptotic form of the radial wavefunction in the $^1S_0$ channel is
\begin{equation}
  u_p(r) = r R(r) \to \tfrac1p \sin[pr + \delta(p)],
\end{equation}
This normalization recovers $R(r) = j_0(r)$ as free solution, such that the full plane wave is normalized as $\phi_{\vec{p}}(\vec{r}) = \exp(i\vec{p}\cdot\vec{r})$.
To be consistent with Refs.~\cite{Cirigliano2020,Cirigliano2021}, we omit the Coulomb interaction from all two-body calculations.

With the wavefunctions obtained from the $R$-matrix formalism, we compute the long- and short-range parts of the amplitude,
\begin{align}
  \mathcal{A}_{L}(p,p') &= 4\pi \int_0^{\infty}\mathrm{d}r \, u_{p'}(r) [V_F(r) + V_{GT}(r)] u_{p}(r) \\
  \mathcal{A}_{S}(p,p') &= 4\pi \int_0^{\infty}\mathrm{d}r' r' \int_0^{\infty}\mathrm{d}r \, r u_{p'}(r') V_S(r,r') u_{p}(r),
\shortintertext{such that}
  \mathcal{A}(p,p') &= \mathcal{A}_{L}(p,p') - 2g \mathcal{A}_{S}(p,p').
\end{align}
We obtain the value of the LEC $g$ by requiring that the total amplitude matches the synthetic datum
\begin{equation}
\mathcal{A}(p=\SI{25}{\MeVc},p'=\SI{30}{\MeVc}) = \SI{-0.0195(5)}{\per\MeV\squared}
\end{equation}
given by Refs.~\cite{Cirigliano2020,Cirigliano2021}.
We validate our calculation of the amplitudes and extraction of the LEC against the results shown in Ref.~\cite{Cirigliano2021}.
See the supplemental material \cite{SupplementalMaterial} for details.

\paragraph{Nucleon-nucleon interactions.}
For the purpose of this work, we employ three different interactions, all derived from chiral effective field theory.
First, we investigate the effect of an SRG transformation on the transition amplitude, employing the N\textsuperscript3LO interaction by \citet{Entem2003}, which we denote by ``EM''.
Next, we perform an analysis of the convergence behavior of the amplitude with respect to the chiral order of the interaction.
For this, we use the family of interactions from \citet{Entem2017}, called ``EMN'' in the following, which provides interactions from LO to N\textsuperscript4LO.
Finally, we consider the $\Delta$N\textsuperscript2LO$_\text{GO}(394)$~\cite{Jiang2020} Hamiltonian, a low-cutoff NN+3N interaction whose construction accounts for $\Delta$ isobars and whose parameters are constrained by $A\leq4$ few-body data as well as nuclear matter properties.
With these interactions, we make the connection to the \emph{ab initio} calculations of the $0\nu\beta\beta$ NME in light nuclei~\cite{Yao2021} and the candidate \nuclide[48]{Ca}~\cite{Yao2020,Belley2021}.

\paragraph{SRG scale dependence.}
In order to accelerate convergence of many-body calculations, the nuclear Hamiltonian is usually preprocessed via unitary transformations that reduce the coupling between low and high momenta.
One choice is the similarity renormalization group \cite{Gazek1993,Wegner1994,Bogner2007}.
The continuous unitary SRG transformation introduces a scale $\lambda$ to the Hamiltonian that controls its bandwidth in momentum space.
The transformation preserves the eigenvalues of $H$ but changes its eigenstates.
Thus, all other observables in principle have to be subject to the same transformation.

\begin{figure}
  \centering
  \includegraphics{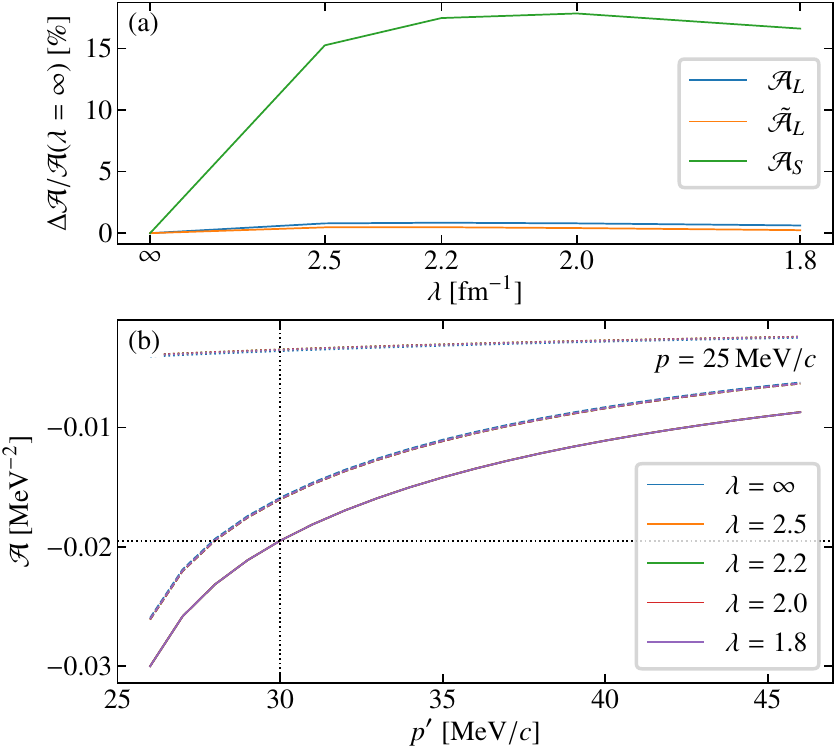}
  \caption{\label{fig:srg dependence}%
    (a) Dependence of the short- and long-range parts of the amplitude on the SRG scale $\lambda$ at the kinematic point $p=\SI{25}{\MeVc}, p'=\SI{30}{\MeVc}$ for the EM potential.
    Shown are the changes relative to the unevolved potential.
    (b) Momentum dependence of the short- and LO long-range parts, as well as the total amplitude for the EM potential at different SRG scales $\lambda$.
    Shown are the scaled short-range part $-2g\mathcal{A}_{S}$ (dotted lines), the long-range part $\mathcal{A}_{L}$ (dashed lines), and the total amplitude $\mathcal{A}_{L} - 2g\mathcal{A}_{S}$ (solid lines).
    The dotted black lines mark the synthetic datum.
    The variation of the total amplitude with respect to $\lambda$ over the momentum range shown is less than \SI{0.1}{\percent}.
  }
\end{figure}

Instead of evolving the $0\nu\beta\beta$ operator exactly, we try to absorb the effect of the SRG by readjusting the contact LEC.
To this end, we calculate the long and short-range amplitudes at the kinematic point using wavefunctions of the Entem and Machleidt interaction at different SRG scales.%
\footnote{Consistent with the regulator for the LO part of the EM interaction, we set $n_\text{exp}=3$ when regularizing the contact.}

The results are shown in \cref{fig:srg dependence}(a).
The long-range part of the amplitude (with or without higher-order corrections) shows a very mild dependence on the SRG scale while the short-range part initially changes by \SI{18}{\percent}.
The change at lower SRG scales is smaller.
This confirms the intuition that the SRG mainly affects short-range operators.

The total amplitudes adjusted to the synthetic datum also change by less than a percent over the range of flow parameters shown.
Overall, the short-range operator enhances the transition amplitude by approximately \SI{22}{\percent} at the kinematic point.
The similar momentum dependence, shown in \cref{fig:srg dependence}(b), implies that the short-range amplitude just acquires a scale-dependent factor $Z(\lambda)$ during the SRG evolution, $\mathcal{A}_{S}(\lambda) = Z(\lambda)\mathcal{A}_{S}(\lambda=\infty)$.
This scaling factor can be compensated by a change in the LEC, resulting in a total amplitude that is virtually independent of the SRG scale once the LEC has been fixed to the synthetic datum.

\begin{table}
  \centering
    \caption{\label{tab:lecs}%
    Value of the $\znubb$ contact LECs for the interactions used in this paper.
    The contact term is regularized using $\Lambda=\SI{500}{\MeVc}$ and $n_\text{exp}=3$ [$\Lambda=\SI{394}{\MeVc}$ and $n_\text{exp}=4$ for $\Delta$N\textsuperscript2LO$_\text{GO}(394)$].
    Amplitudes are shown in units of \si{\MeV\tothe{-2}} at the kinematic point $p=\SI{25}{\MeVc}$, $p'=\SI{30}{\MeVc}$.
    The quantities with a tilde incorporate beyond-LO effects in the operator.
    The quoted uncertainties $\Delta g$ are propagated from the uncertainty of the synthetic datum and are identical for $g$ and $\tilde{g}$.
    See supplemental material \cite{SupplementalMaterial} for recommended values at other SRG scales and chiral orders.
  }
  \begin{tabular}{lccccccc}
  \toprule
Interaction & $\lambda$ & $10^3 \tilde{\mathcal{A}}_L$ & $10^3 \mathcal{A}_L$ & $10^3 \mathcal{A}_S$ & $\tilde{g}$ & $g$ & $\Delta g$ \\
  \midrule
EM & $\infty$ & $-15.847$ & $-15.898$ & $3.0152$ & $0.606$ & $0.597$ & $0.083$ \\
   & $2.50$ & $-15.921$ & $-16.024$ & $3.0635$ & $0.584$ & $0.567$ & $0.082$ \\
   & $2.24$ & $-15.923$ & $-16.033$ & $3.0451$ & $0.587$ & $0.569$ & $0.082$ \\
   & $2.20$ & $-15.923$ & $-16.033$ & $3.0408$ & $0.588$ & $0.570$ & $0.082$ \\
   & $2.00$ & $-15.912$ & $-16.025$ & $3.0061$ & $0.597$ & $0.578$ & $0.083$ \\
   & $1.88$ & $-15.898$ & $-16.011$ & $2.9733$ & $0.606$ & $0.587$ & $0.084$ \\
   & $1.80$ & $-15.885$ & $-15.998$ & $2.9446$ & $0.614$ & $0.595$ & $0.085$ \\
\midrule
EMN N\textsuperscript3LO & $\infty$ & $-15.857$ & $-15.903$ & $2.3816$ & $0.765$ & $0.755$ & $0.105$ \\
    & $2.00$   & $-15.934$ & $-16.043$ & $2.9031$ & $0.614$ & $0.595$ & $0.086$ \\
\midrule
$\Delta$N\textsuperscript2LO$_\text{GO}$ & $\infty$ & $-15.846$ & $-15.968$ & $3.1225$ & $0.585$ & $0.566$ & $0.080$ \\
(394)   & $2.00$ & $-15.776$ & $-15.892$ & $2.9610$ & $0.629$ & $0.609$ & $0.084$ \\
\bottomrule
  \end{tabular}
\end{table}

\paragraph{Convergence of the chiral expansion.}
Next, we consider the dependence of the $\znubb$ amplitude on the order of the chiral interaction employed.
To this end we use the EMN family of interactions from LO up to N\textsuperscript4LO~\cite{Entem2017}.

\begin{figure}
  \centering
  \includegraphics{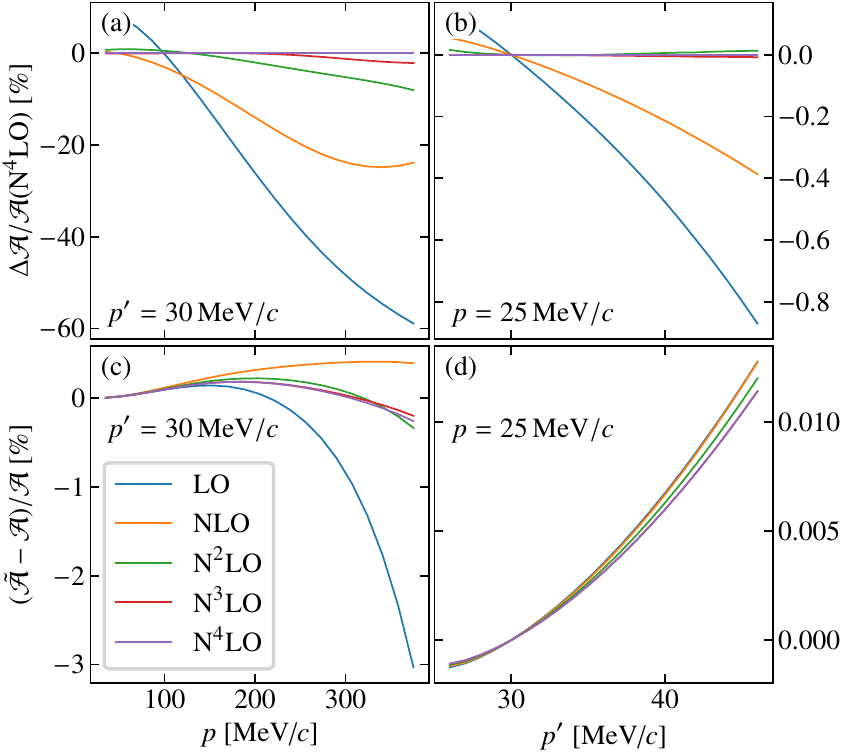}
  \caption{\label{fig:order ratio momentum}%
    (a-b) Ratio of total amplitudes (relative to the N\textsuperscript4LO result) for different orders of the chiral expansion as a function of incoming and outgoing momentum, respectively.
    (c-d) Relative difference between the amplitudes using the LO operator and the operator containing beyond-LO corrections as a function of incoming and outgoing momentum, respectively.
  }
\end{figure}

We consider the full amplitude as a function of incoming and outgoing relative momenta for different chiral orders.
For incoming momenta up to \SI{375}{\MeVc}, the range up to which the potentials are fitted, we notice a sizable dependence on the chiral order, which is shown in \cref{fig:order ratio momentum}(a).
The total amplitude computed with the LO interaction drops by more than \SI{60}{\percent} compared to N\textsuperscript4LO, but systematically converges to the N\textsuperscript4LO result with increasing order.
The variation in the low-momentum region [cf.\ \cref{fig:order ratio momentum}(b)] is less than \SI{1}{\percent} and also rapidly converging.

Finally, we investigate the effect of including beyond-LO terms into the $\znubb$ operator by employing the neutrino potentials from \cref{eq:neutrino pheno}.
The phenomenological corrections added there only modify the potential at short range.
At distances $r > \SI{1.5}{\fm}$ they are virtually indistinguishable from the LO ones.
Since we use NN interactions with relatively low cutoffs, the total amplitudes are fairly insensitive to the short-range modifications.
The relative difference between them is below \SI{0.5}{\percent} for momenta within the range of applicability of the respective NN interaction.
For the LO NN interaction the difference may reach \SI{3}{\percent} at incoming momenta exceeding \SI{300}{\MeVc} [cf.\ \cref{fig:order ratio momentum}(c)].
The difference between both amplitudes at low momenta, shown in \cref{fig:order ratio momentum}(d), is negligible.
\Cref{tab:lecs} shows the long-range amplitudes and LECs $\mathcal{A}$, $g$ and $\tilde{\mathcal{A}}$, $\tilde{g}$ associated with the LO long-range operator and its extension, respectively.

\paragraph{Application to finite nuclei.}
While suitable for generating the synthetic datum, a scattering state of neutrons is not ideal for observing $\znubb$ decay in experiment.
For that, we need to move to finite nuclei for which the single-$\beta$ decay is energetically forbidden.
Due to the long lifetime any competing decay would drown out the $\znubb$-decay signal.
A few candidate nuclei that fulfill this requirement have been identified, some of which can even be used to build an active detector.

Previous calculations of the NME in finite nuclei only considered the long-range part of the operator.
With the LEC of the short-range part of the operator adjusted to the synthetic datum, we can now calculate its effect and provide a first result renormalized to leading order.
Here, we revisit our benchmark calculations for light nuclei~\cite{Yao2021}, as well as the candidate pair \nuclide[48]{Ca} and \nuclide[48]{Ti} \cite{Yao2020}.
The interaction used in these studies is the so-called EM1.8/2.0 \cite{Hebeler2011}, which consists of the EM interaction SRG-evolved to a scale $\lambda=\SI{1.8}{\per\fm}$ augmented by an unevolved N\textsuperscript2LO three-nucleon interaction.
To estimate the dependence of the NME on SRG scale and chiral order, we additionally consider Hamiltonians based on the EM interaction with a local-nonlocal 3N force~\cite{Soma2020}, called ``LNL'' here, one that combines the EMN N\textsuperscript3LO with an N\textsuperscript2LO 3N interaction \cite{Huther2020} (designated there as N\textsuperscript3LO'), and the $\Delta$N\textsuperscript2LO$_\text{GO}(394)$ NN+3N Hamiltonian.
The LECs for each of the NN interactions are shown in \cref{tab:lecs}.

The NME for finite nuclei is defined as
\begin{equation}
  M^{0\nu} = \frac{4\pi R}{g_A^2} \braket{\nuclide[A]{(Z+2)}|\hat{\tilde{V}}_{F}+\hat{\tilde{V}}_{GT}+\hat{\tilde{V}}_{T} - 2\tilde g\hat{V}_{S}|\nuclide[A]{Z}},
\end{equation}
with the empirical nuclear radius $R=R_0 A^{1/3}$ and $R_0 = \SI{1.2}{\fm}$.
The operator $\hat{\tilde{V}}_{T}$ contains the tensor part of the decay operator.
With this definition, $M^{0\nu}$ is dimensionless.

\begin{figure}
  \centering
  \includegraphics[width=0.9\columnwidth]{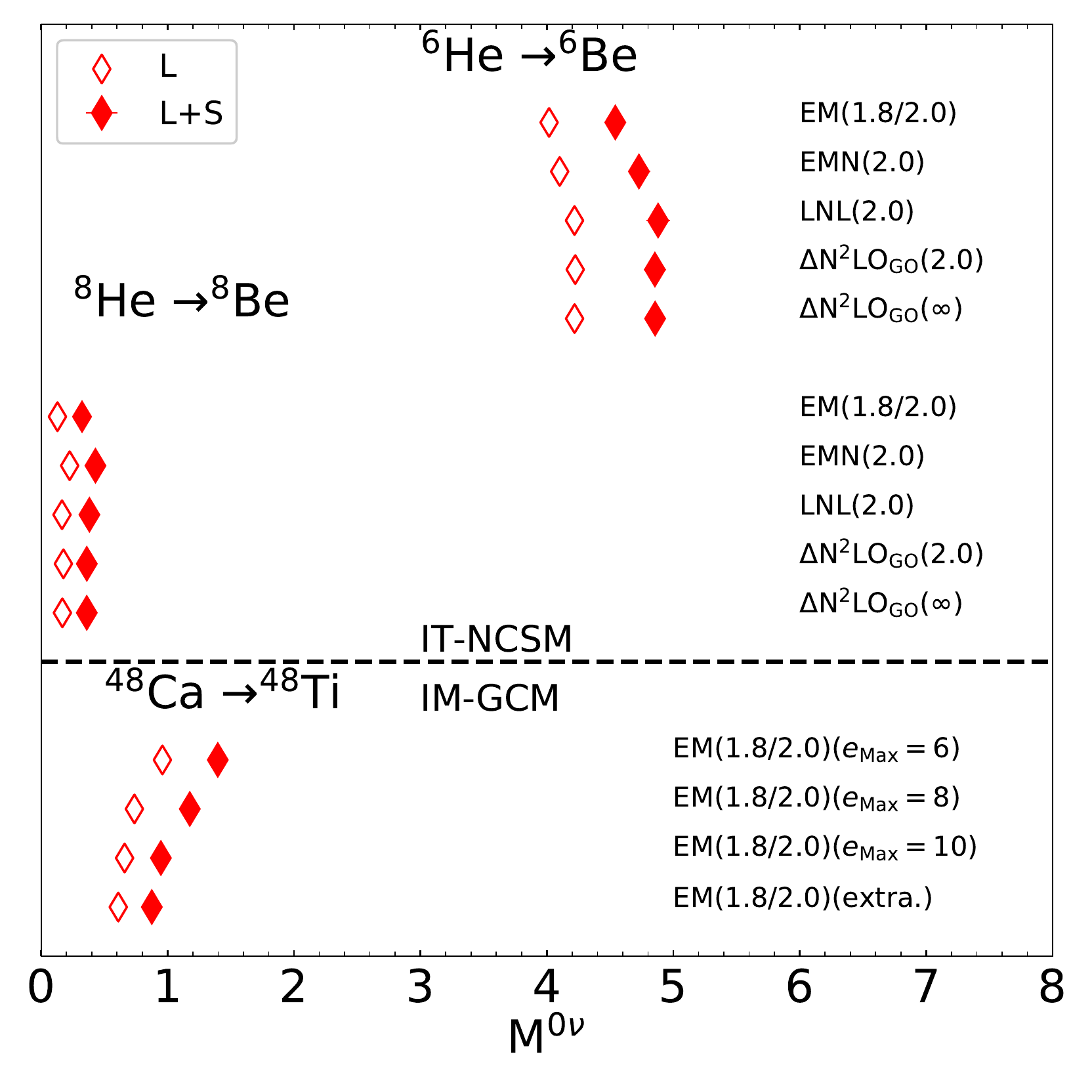}
  \caption{\label{fig:NME4ALL}%
   The NMEs $M^{0\nu}$ of  isospin-conserving ($\Delta T=0$) transition \nuclide[6]{He} $\to$ \nuclide[6]{Be}, and isospin-nonconserving  ($\Delta T=2$) transitions \nuclide[8]{He} $\to$ \nuclide[8]{Be} and \nuclide[48]{Ca} $\to$ \nuclide[48]{Ti}, calculated with different chiral nuclear forces and with both long- and short-range transition operators.
  }
\end{figure}

First, we investigate the NME in the pairs of light nuclei \nuclide[6]{He}--\nuclide[6]{Be} and \nuclide[8]{He}--\nuclide[8]{Be} as examples of $\Delta T=0$ and $\Delta T=2$ transitions with the importance-truncated no-core shell model (IT-NCSM) \cite{Roth2009}.
The results are summarized in \cref{fig:NME4ALL}.
We note that the contact operator increases the NME by a factor ranging from \SI{11}{\percent} to \SI{17}{\percent} for $\Delta T=0$ transition in \nuclide[6]{He}.
Transitions with $\Delta T=2$ have a node in the transition density that leads to a cancellation between short and long distances.
This cancellation affects the long-range part more strongly than the contact, leading to small overall NMEs and relatively larger contributions of the contact term.
Thus, the contact increases the $\Delta T=2$ transition in \nuclide[8]{He} by \SI{92}{\percent} to \SI{172}{\percent}.
Overall, SRG-transforming the $\Delta$N\textsuperscript2LO$_\text{GO}$ as well as switching to the LNL Hamiltonian barely changes the NME.
Despite using the same NN interaction at a similar SRG scale as the LNL, the EM1.8/2.0 produces systematically smaller NMEs than the other interactions.
The EMN + N\textsuperscript3LO' Hamiltonian yields a smaller NME in \nuclide[6]{He} than the LNL while the \nuclide[8]{He} NME is larger.
Both are driven by the long-range part, the short-range contribution is of similar size compared to the LNL Hamiltonian.
This shows that there is still some uncertainty stemming from the Hamiltonian, in particular the 3N interaction, which needs to be quantified further.

For the lightest $\znubb$-decay candidate nucleus \nuclide[48]{Ca}, the short-range operator increases the NME by \SI{43(7)}{\percent}.
With this contribution, the value of $M^{0\nu}$ is \num{0.875(40)} for \nuclide[48]{Ca} from the in-medium generator coordinate method (IM-GCM) \cite{Yao2020} calculation, the uncertainty of which is from the LEC $\tilde g$ of the short-range transition operator.

\paragraph{Conclusions and outlook.}
In this work, we present a determination of the LEC of a contact operator that enters the $\znubb$ operator at leading order for a set of chiral interactions, which are used in \emph{ab initio} calculations of nuclei.
To fix the LEC, we take the synthetic datum provided by Cirigliano \emph{et al.}~\cite{Cirigliano2020,Cirigliano2021}, which contains the effect of light Majorana-neutrino exchange.
We investigate the dependence of the $nn\to ppe^-e^-$ amplitude on the SRG scale and order of the interaction.
We find that a change in the SRG scale can be compensated by readjusting the LEC, leading to only very small changes in the total amplitude.
The dependence on the order of the interaction can be sizable for low-order interactions at high momenta beyond the range applicability of the respective interaction.
However, the total amplitude converges quickly when going beyond N\textsuperscript2LO over the full momentum range to which the potential is fitted.
The robustness of the amplitude shows that the two-body system is under control and any changes in the momentum dependence will come from subleading terms in the operator.
Moreover, the changes will likely be small because including beyond-LO terms in the long-range part barely changes the result, apart from a change in the LEC.

The contact operator turns out to increase significantly the NME of isospin-changing transition in finite nuclei. For the lightest candidate nucleus  $\nuclide[48]{Ca}$, the NME is enhanced by \SI{43(7)}{\percent}.
This enhancement is also found in the ab initio calculations of light nuclei $\nuclide[6,8]{He}$ using the  three families of chiral interactions with a low-scale regulator.
It indicates that the contact operator will generally enhance the NMEs predicted by ab initio many-body calculations using these interactions and this effect should be taken into account in the future ab initio calculations.
The extension of current studies to the NMEs of $\znubb$ decays in heavier candidate nuclei is highly interesting.

We note that the present work relies on the synthetic datum, the uncertainty of which is dominated by neglected inelastic contributions, and hopefully is to be reduced in a future lattice QCD calculation~\cite{Davoudi:2021}.
Nevertheless, apart from the total NMEs all the findings presented here are independent of the concrete value of the synthetic datum.
The availability of a more precise datum will just result in a shift of the total amplitudes, and we provide separate short- and long-range parts to enable matching to an updated value.

\begin{acknowledgments}
We thank V.~Cirigliano, W.~Dekens, J.~de Vries, J.~Engel, L.S.~Geng, M.~Hoferichter, B. W.~Long and E.~Mereghetti for fruitful discussions, A.~Ekstr\"om and R.~Machleidt for providing us with their NN interaction routines, as well as K.~Hebeler for providing momentum-space three-nucleon interaction matrix elements.
This work is supported in part by the U.S.\ Department of Energy, Office of Science, Office of Nuclear Physics under Awards No.\ DE-SC0017887, No.\ DE-SC0015376 (the DBD Topical Theory Collaboration), and No.\ DE-SC0018083 (NUCLEI SciDAC-4 Collaboration).
Computing resources were provided by the Institute for Cyber-Enabled Research at Michigan State University, and the U.S.\ National Energy Research Scientific Computing Center (NERSC), a DOE Office of Science User Facility supported by the Office of Science of the U.S.\ Department of Energy under Contract No.\ DE-AC02-05CH11231.
\end{acknowledgments}

\bibliography{abbrev,lib,extra}

\begin{thebibliography}{43}%
\makeatletter
\providecommand \@ifxundefined [1]{%
 \@ifx{#1\undefined}
}%
\providecommand \@ifnum [1]{%
 \ifnum #1\expandafter \@firstoftwo
 \else \expandafter \@secondoftwo
 \fi
}%
\providecommand \@ifx [1]{%
 \ifx #1\expandafter \@firstoftwo
 \else \expandafter \@secondoftwo
 \fi
}%
\providecommand \natexlab [1]{#1}%
\providecommand \enquote  [1]{``#1''}%
\providecommand \bibnamefont  [1]{#1}%
\providecommand \bibfnamefont [1]{#1}%
\providecommand \citenamefont [1]{#1}%
\providecommand \href@noop [0]{\@secondoftwo}%
\providecommand \href [0]{\begingroup \@sanitize@url \@href}%
\providecommand \@href[1]{\@@startlink{#1}\@@href}%
\providecommand \@@href[1]{\endgroup#1\@@endlink}%
\providecommand \@sanitize@url [0]{\catcode `\\12\catcode `\$12\catcode
  `\&12\catcode `\#12\catcode `\^12\catcode `\_12\catcode `\%12\relax}%
\providecommand \@@startlink[1]{}%
\providecommand \@@endlink[0]{}%
\providecommand \url  [0]{\begingroup\@sanitize@url \@url }%
\providecommand \@url [1]{\endgroup\@href {#1}{\urlprefix }}%
\providecommand \urlprefix  [0]{URL }%
\providecommand \Eprint [0]{\href }%
\providecommand \doibase [0]{https://doi.org/}%
\providecommand \selectlanguage [0]{\@gobble}%
\providecommand \bibinfo  [0]{\@secondoftwo}%
\providecommand \bibfield  [0]{\@secondoftwo}%
\providecommand \translation [1]{[#1]}%
\providecommand \BibitemOpen [0]{}%
\providecommand \bibitemStop [0]{}%
\providecommand \bibitemNoStop [0]{.\EOS\space}%
\providecommand \EOS [0]{\spacefactor3000\relax}%
\providecommand \BibitemShut  [1]{\csname bibitem#1\endcsname}%
\let\auto@bib@innerbib\@empty
\bibitem [{\citenamefont {Schechter}\ and\ \citenamefont
  {Valle}(1982)}]{Schechter:1982}%
  \BibitemOpen
  \bibfield  {author} {\bibinfo {author} {\bibfnamefont {J.}~\bibnamefont
  {Schechter}}\ and\ \bibinfo {author} {\bibfnamefont {J.~W.~F.}\ \bibnamefont
  {Valle}},\ }\bibfield  {title} {\bibinfo {title} {{Neutrinoless
  double-$\ensuremath{\beta}$ decay in $\mathrm{SU}(2)\times \mathrm{U}(1)$
  theories}},\ }\href {https://doi.org/10.1103/PhysRevD.25.2951} {\bibfield
  {journal} {\bibinfo  {journal} {Phys. Rev. D}\ }\textbf {\bibinfo {volume}
  {25}},\ \bibinfo {pages} {2951} (\bibinfo {year} {1982})}\BibitemShut
  {NoStop}%
\bibitem [{\citenamefont {Men\'endez}\ \emph {et~al.}(2009)\citenamefont
  {Men\'endez}, \citenamefont {Poves}, \citenamefont {Caurier},\ and\
  \citenamefont {Nowacki}}]{Menendez:2009}%
  \BibitemOpen
  \bibfield  {author} {\bibinfo {author} {\bibfnamefont {J.}~\bibnamefont
  {Men\'endez}}, \bibinfo {author} {\bibfnamefont {A.}~\bibnamefont {Poves}},
  \bibinfo {author} {\bibfnamefont {E.}~\bibnamefont {Caurier}},\ and\ \bibinfo
  {author} {\bibfnamefont {F.}~\bibnamefont {Nowacki}},\ }\bibfield  {title}
  {\bibinfo {title} {Disassembling the nuclear matrix elements of the
  neutrinoless $\ensuremath{\beta}\ensuremath{\beta}$ decay},\ }\href
  {https://doi.org/10.1016/j.nuclphysa.2008.12.005} {\bibfield  {journal}
  {\bibinfo  {journal} {Nucl. Phys. A}\ }\textbf {\bibinfo {volume} {818}},\
  \bibinfo {pages} {139 } (\bibinfo {year} {2009})}\BibitemShut {NoStop}%
\bibitem [{\citenamefont {Rodr\'{\i}guez}\ and\ \citenamefont
  {Mart\'{\i}nez-Pinedo}(2010)}]{Rodriguez:2010}%
  \BibitemOpen
  \bibfield  {author} {\bibinfo {author} {\bibfnamefont {T.~R.}\ \bibnamefont
  {Rodr\'{\i}guez}}\ and\ \bibinfo {author} {\bibfnamefont {G.}~\bibnamefont
  {Mart\'{\i}nez-Pinedo}},\ }\bibfield  {title} {\bibinfo {title} {Energy
  density functional study of nuclear matrix elements for neutrinoless
  $\ensuremath{\beta}\ensuremath{\beta}$ decay},\ }\href
  {https://doi.org/10.1103/PhysRevLett.105.252503} {\bibfield  {journal}
  {\bibinfo  {journal} {Phys. Rev. Lett.}\ }\textbf {\bibinfo {volume} {105}},\
  \bibinfo {pages} {252503} (\bibinfo {year} {2010})}\BibitemShut {NoStop}%
\bibitem [{\citenamefont {Barea}\ \emph {et~al.}(2013)\citenamefont {Barea},
  \citenamefont {Kotila},\ and\ \citenamefont {Iachello}}]{Barea:2013}%
  \BibitemOpen
  \bibfield  {author} {\bibinfo {author} {\bibfnamefont {J.}~\bibnamefont
  {Barea}}, \bibinfo {author} {\bibfnamefont {J.}~\bibnamefont {Kotila}},\ and\
  \bibinfo {author} {\bibfnamefont {F.}~\bibnamefont {Iachello}},\ }\bibfield
  {title} {\bibinfo {title} {Nuclear matrix elements for
  double-$\ensuremath{\beta}$ decay},\ }\href
  {https://doi.org/10.1103/PhysRevC.87.014315} {\bibfield  {journal} {\bibinfo
  {journal} {Phys. Rev. C}\ }\textbf {\bibinfo {volume} {87}},\ \bibinfo
  {pages} {014315} (\bibinfo {year} {2013})}\BibitemShut {NoStop}%
\bibitem [{\citenamefont {Mustonen}\ and\ \citenamefont
  {Engel}(2013)}]{Mustonen:2013}%
  \BibitemOpen
  \bibfield  {author} {\bibinfo {author} {\bibfnamefont {M.~T.}\ \bibnamefont
  {Mustonen}}\ and\ \bibinfo {author} {\bibfnamefont {J.}~\bibnamefont
  {Engel}},\ }\bibfield  {title} {\bibinfo {title} {{Large-scale calculations
  of the double-$\beta$ decay of $^{76}$Ge, $^{130}$Te, $^{136}$Xe, and
  $^{150}$Nd in the deformed self-consistent Skyrme quasiparticle random-phase
  approximation}},\ }\href {https://doi.org/10.1103/PhysRevC.87.064302}
  {\bibfield  {journal} {\bibinfo  {journal} {Phys. Rev. C}\ }\textbf {\bibinfo
  {volume} {87}},\ \bibinfo {pages} {064302} (\bibinfo {year}
  {2013})}\BibitemShut {NoStop}%
\bibitem [{\citenamefont {Holt}\ and\ \citenamefont {Engel}(2013)}]{Holt:2013}%
  \BibitemOpen
  \bibfield  {author} {\bibinfo {author} {\bibfnamefont {J.~D.}\ \bibnamefont
  {Holt}}\ and\ \bibinfo {author} {\bibfnamefont {J.}~\bibnamefont {Engel}},\
  }\bibfield  {title} {\bibinfo {title} {{Effective
  double-\ensuremath{\beta}-decay operator for $^{76}$Ge and $^{82}$Se}},\
  }\href {https://doi.org/10.1103/PhysRevC.87.064315} {\bibfield  {journal}
  {\bibinfo  {journal} {Phys. Rev. C}\ }\textbf {\bibinfo {volume} {87}},\
  \bibinfo {pages} {064315} (\bibinfo {year} {2013})}\BibitemShut {NoStop}%
\bibitem [{\citenamefont {Kwiatkowski}\ \emph {et~al.}(2014)\citenamefont
  {Kwiatkowski}, \citenamefont {Brunner}, \citenamefont {Holt}, \citenamefont
  {Chaudhuri}, \citenamefont {Chowdhury}, \citenamefont {Eibach}, \citenamefont
  {Engel}, \citenamefont {Gallant}, \citenamefont {Grossheim}, \citenamefont
  {Horoi}, \citenamefont {Lennarz}, \citenamefont {Macdonald}, \citenamefont
  {Pearson}, \citenamefont {Schultz}, \citenamefont {Simon}, \citenamefont
  {Senkov}, \citenamefont {Simon}, \citenamefont {Zuber},\ and\ \citenamefont
  {Dilling}}]{Kwiatkowski:2014}%
  \BibitemOpen
  \bibfield  {author} {\bibinfo {author} {\bibfnamefont {A.~A.}\ \bibnamefont
  {Kwiatkowski}}, \bibinfo {author} {\bibfnamefont {T.}~\bibnamefont
  {Brunner}}, \bibinfo {author} {\bibfnamefont {J.~D.}\ \bibnamefont {Holt}},
  \bibinfo {author} {\bibfnamefont {A.}~\bibnamefont {Chaudhuri}}, \bibinfo
  {author} {\bibfnamefont {U.}~\bibnamefont {Chowdhury}}, \bibinfo {author}
  {\bibfnamefont {M.}~\bibnamefont {Eibach}}, \bibinfo {author} {\bibfnamefont
  {J.}~\bibnamefont {Engel}}, \bibinfo {author} {\bibfnamefont {A.~T.}\
  \bibnamefont {Gallant}}, \bibinfo {author} {\bibfnamefont {A.}~\bibnamefont
  {Grossheim}}, \bibinfo {author} {\bibfnamefont {M.}~\bibnamefont {Horoi}},
  \bibinfo {author} {\bibfnamefont {A.}~\bibnamefont {Lennarz}}, \bibinfo
  {author} {\bibfnamefont {T.~D.}\ \bibnamefont {Macdonald}}, \bibinfo {author}
  {\bibfnamefont {M.~R.}\ \bibnamefont {Pearson}}, \bibinfo {author}
  {\bibfnamefont {B.~E.}\ \bibnamefont {Schultz}}, \bibinfo {author}
  {\bibfnamefont {M.~C.}\ \bibnamefont {Simon}}, \bibinfo {author}
  {\bibfnamefont {R.~A.}\ \bibnamefont {Senkov}}, \bibinfo {author}
  {\bibfnamefont {V.~V.}\ \bibnamefont {Simon}}, \bibinfo {author}
  {\bibfnamefont {K.}~\bibnamefont {Zuber}},\ and\ \bibinfo {author}
  {\bibfnamefont {J.}~\bibnamefont {Dilling}},\ }\bibfield  {title} {\bibinfo
  {title} {New determination of double-$\ensuremath{\beta}$-decay properties in
  ${}^{48}\mathrm{Ca}$: High-precision
  ${Q}_{\ensuremath{\beta}\ensuremath{\beta}}$-value measurement and improved
  nuclear matrix element calculations},\ }\href
  {https://doi.org/10.1103/PhysRevC.89.045502} {\bibfield  {journal} {\bibinfo
  {journal} {Phys. Rev. C}\ }\textbf {\bibinfo {volume} {89}},\ \bibinfo
  {pages} {045502} (\bibinfo {year} {2014})}\BibitemShut {NoStop}%
\bibitem [{\citenamefont {Song}\ \emph {et~al.}(2014)\citenamefont {Song},
  \citenamefont {Yao}, \citenamefont {Ring},\ and\ \citenamefont
  {Meng}}]{Song:2014}%
  \BibitemOpen
  \bibfield  {author} {\bibinfo {author} {\bibfnamefont {L.~S.}\ \bibnamefont
  {Song}}, \bibinfo {author} {\bibfnamefont {J.~M.}\ \bibnamefont {Yao}},
  \bibinfo {author} {\bibfnamefont {P.}~\bibnamefont {Ring}},\ and\ \bibinfo
  {author} {\bibfnamefont {J.}~\bibnamefont {Meng}},\ }\bibfield  {title}
  {\bibinfo {title} {Relativistic description of nuclear matrix elements in
  neutrinoless double-$\ensuremath{\beta}$ decay},\ }\href
  {https://doi.org/https://doi.org/10.1103/PhysRevC.90.054309} {\bibfield
  {journal} {\bibinfo  {journal} {Phys. Rev. C}\ }\textbf {\bibinfo {volume}
  {90}},\ \bibinfo {pages} {054309} (\bibinfo {year} {2014})}\BibitemShut
  {NoStop}%
\bibitem [{\citenamefont {Yao}\ \emph {et~al.}(2015)\citenamefont {Yao},
  \citenamefont {Song}, \citenamefont {Hagino}, \citenamefont {Ring},\ and\
  \citenamefont {Meng}}]{Yao:2015}%
  \BibitemOpen
  \bibfield  {author} {\bibinfo {author} {\bibfnamefont {J.~M.}\ \bibnamefont
  {Yao}}, \bibinfo {author} {\bibfnamefont {L.~S.}\ \bibnamefont {Song}},
  \bibinfo {author} {\bibfnamefont {K.}~\bibnamefont {Hagino}}, \bibinfo
  {author} {\bibfnamefont {P.}~\bibnamefont {Ring}},\ and\ \bibinfo {author}
  {\bibfnamefont {J.}~\bibnamefont {Meng}},\ }\bibfield  {title} {\bibinfo
  {title} {Systematic study of nuclear matrix elements in neutrinoless
  double-$\ensuremath{\beta}$ decay with a beyond-mean-field covariant density
  functional theory},\ }\href {https://doi.org/10.1103/PhysRevC.91.024316}
  {\bibfield  {journal} {\bibinfo  {journal} {Phys. Rev. C}\ }\textbf {\bibinfo
  {volume} {91}},\ \bibinfo {pages} {024316} (\bibinfo {year}
  {2015})}\BibitemShut {NoStop}%
\bibitem [{\citenamefont {Hyv\"arinen}\ and\ \citenamefont
  {Suhonen}(2015)}]{Hyvarinen:2015}%
  \BibitemOpen
  \bibfield  {author} {\bibinfo {author} {\bibfnamefont {J.}~\bibnamefont
  {Hyv\"arinen}}\ and\ \bibinfo {author} {\bibfnamefont {J.}~\bibnamefont
  {Suhonen}},\ }\bibfield  {title} {\bibinfo {title} {Nuclear matrix elements
  for $0\ensuremath{\nu}\ensuremath{\beta}\ensuremath{\beta}$ decays with light
  or heavy majorana-neutrino exchange},\ }\href
  {https://doi.org/10.1103/PhysRevC.91.024613} {\bibfield  {journal} {\bibinfo
  {journal} {Phys. Rev. C}\ }\textbf {\bibinfo {volume} {91}},\ \bibinfo
  {pages} {024613} (\bibinfo {year} {2015})}\BibitemShut {NoStop}%
\bibitem [{\citenamefont {Horoi}\ and\ \citenamefont
  {Neacsu}(2016)}]{Horoi:2016}%
  \BibitemOpen
  \bibfield  {author} {\bibinfo {author} {\bibfnamefont {M.}~\bibnamefont
  {Horoi}}\ and\ \bibinfo {author} {\bibfnamefont {A.}~\bibnamefont {Neacsu}},\
  }\bibfield  {title} {\bibinfo {title} {Shell model predictions for
  $^{124}\mathrm{Sn}$ double-$\ensuremath{\beta}$ decay},\ }\href
  {https://doi.org/10.1103/PhysRevC.93.024308} {\bibfield  {journal} {\bibinfo
  {journal} {Phys. Rev. C}\ }\textbf {\bibinfo {volume} {93}},\ \bibinfo
  {pages} {024308} (\bibinfo {year} {2016})}\BibitemShut {NoStop}%
\bibitem [{\citenamefont {Song}\ \emph {et~al.}(2017)\citenamefont {Song},
  \citenamefont {Yao}, \citenamefont {Ring},\ and\ \citenamefont
  {Meng}}]{Song:2017}%
  \BibitemOpen
  \bibfield  {author} {\bibinfo {author} {\bibfnamefont {L.~S.}\ \bibnamefont
  {Song}}, \bibinfo {author} {\bibfnamefont {J.~M.}\ \bibnamefont {Yao}},
  \bibinfo {author} {\bibfnamefont {P.}~\bibnamefont {Ring}},\ and\ \bibinfo
  {author} {\bibfnamefont {J.}~\bibnamefont {Meng}},\ }\bibfield  {title}
  {\bibinfo {title} {Nuclear matrix element of neutrinoless
  double-$\ensuremath{\beta}$ decay: Relativity and short-range correlations},\
  }\href {https://doi.org/10.1103/PhysRevC.95.024305} {\bibfield  {journal}
  {\bibinfo  {journal} {Phys. Rev. C}\ }\textbf {\bibinfo {volume} {95}},\
  \bibinfo {pages} {024305} (\bibinfo {year} {2017})}\BibitemShut {NoStop}%
\bibitem [{\citenamefont {Jiao}\ \emph {et~al.}(2017)\citenamefont {Jiao},
  \citenamefont {Engel},\ and\ \citenamefont {Holt}}]{Jiao:2017}%
  \BibitemOpen
  \bibfield  {author} {\bibinfo {author} {\bibfnamefont {C.~F.}\ \bibnamefont
  {Jiao}}, \bibinfo {author} {\bibfnamefont {J.}~\bibnamefont {Engel}},\ and\
  \bibinfo {author} {\bibfnamefont {J.~D.}\ \bibnamefont {Holt}},\ }\bibfield
  {title} {\bibinfo {title} {Neutrinoless double-$\ensuremath{\beta}$ decay
  matrix elements in large shell-model spaces with the generator-coordinate
  method},\ }\href {https://doi.org/10.1103/PhysRevC.96.054310} {\bibfield
  {journal} {\bibinfo  {journal} {Phys. Rev. C}\ }\textbf {\bibinfo {volume}
  {96}},\ \bibinfo {pages} {054310} (\bibinfo {year} {2017})}\BibitemShut
  {NoStop}%
\bibitem [{\citenamefont {Yoshinaga}\ \emph {et~al.}(2018)\citenamefont
  {Yoshinaga}, \citenamefont {Yanase}, \citenamefont {Higashiyama},
  \citenamefont {Teruya},\ and\ \citenamefont {Taguchi}}]{Yoshinaga:2018}%
  \BibitemOpen
  \bibfield  {author} {\bibinfo {author} {\bibfnamefont {N.}~\bibnamefont
  {Yoshinaga}}, \bibinfo {author} {\bibfnamefont {K.}~\bibnamefont {Yanase}},
  \bibinfo {author} {\bibfnamefont {K.}~\bibnamefont {Higashiyama}}, \bibinfo
  {author} {\bibfnamefont {E.}~\bibnamefont {Teruya}},\ and\ \bibinfo {author}
  {\bibfnamefont {D.}~\bibnamefont {Taguchi}},\ }\bibfield  {title} {\bibinfo
  {title} {{Structure of nuclei with masses 76 and 82 and nuclear matrix
  elements of neutrinoless double beta decay}},\ }\href
  {https://doi.org/10.1093/ptep/ptx174} {\bibfield  {journal} {\bibinfo
  {journal} {Prog. Theor. Exp. Phys.}\ }\textbf {\bibinfo {volume} {2018}},\
  \bibinfo {pages} {023D02} (\bibinfo {year} {2018})}\BibitemShut {NoStop}%
\bibitem [{\citenamefont {Fang}\ \emph {et~al.}(2018)\citenamefont {Fang},
  \citenamefont {Faessler},\ and\ \citenamefont {\ifmmode~\check{S}\else
  \v{S}\fi{}imkovic}}]{Fang:2018}%
  \BibitemOpen
  \bibfield  {author} {\bibinfo {author} {\bibfnamefont {D.-L.}\ \bibnamefont
  {Fang}}, \bibinfo {author} {\bibfnamefont {A.}~\bibnamefont {Faessler}},\
  and\ \bibinfo {author} {\bibfnamefont {F.}~\bibnamefont
  {\ifmmode~\check{S}\else \v{S}\fi{}imkovic}},\ }\bibfield  {title} {\bibinfo
  {title} {$0\ensuremath{\nu}\ensuremath{\beta}\ensuremath{\beta}$-decay
  nuclear matrix element for light and heavy neutrino mass mechanisms from
  deformed quasiparticle random-phase approximation calculations for
  $^{76}\mathrm{Ge}, ^{82}\mathrm{Se}, ^{130}\mathrm{Te}, ^{136}\mathrm{Xe}$,
  and $^{150}\mathrm{Nd}$ with isospin restoration},\ }\href
  {https://doi.org/10.1103/PhysRevC.97.045503} {\bibfield  {journal} {\bibinfo
  {journal} {Phys. Rev. C}\ }\textbf {\bibinfo {volume} {97}},\ \bibinfo
  {pages} {045503} (\bibinfo {year} {2018})}\BibitemShut {NoStop}%
\bibitem [{\citenamefont {Rath}\ \emph {et~al.}(2019)\citenamefont {Rath},
  \citenamefont {Chandra}, \citenamefont {Chaturvedi},\ and\ \citenamefont
  {Raina}}]{Rath:2019}%
  \BibitemOpen
  \bibfield  {author} {\bibinfo {author} {\bibfnamefont {P.~K.}\ \bibnamefont
  {Rath}}, \bibinfo {author} {\bibfnamefont {R.}~\bibnamefont {Chandra}},
  \bibinfo {author} {\bibfnamefont {K.}~\bibnamefont {Chaturvedi}},\ and\
  \bibinfo {author} {\bibfnamefont {P.~K.}\ \bibnamefont {Raina}},\ }\bibfield
  {title} {\bibinfo {title} {Nuclear transition matrix elements for
  double-{$\beta$} decay within phfb model},\ }\href
  {https://doi.org/10.3389/fphy.2019.00064} {\bibfield  {journal} {\bibinfo
  {journal} {Frontiers in Physics}\ }\textbf {\bibinfo {volume} {7}},\ \bibinfo
  {pages} {64} (\bibinfo {year} {2019})}\BibitemShut {NoStop}%
\bibitem [{\citenamefont {Terasaki}\ and\ \citenamefont
  {Iwata}(2019)}]{Terasaki:2019}%
  \BibitemOpen
  \bibfield  {author} {\bibinfo {author} {\bibfnamefont {J.}~\bibnamefont
  {Terasaki}}\ and\ \bibinfo {author} {\bibfnamefont {Y.}~\bibnamefont
  {Iwata}},\ }\bibfield  {title} {\bibinfo {title} {Isoscalar pairing
  interaction for the quasiparticle random-phase approximation approach to
  double-$\ensuremath{\beta}$ and $\ensuremath{\beta}$ decays},\ }\href
  {https://doi.org/10.1103/PhysRevC.100.034325} {\bibfield  {journal} {\bibinfo
   {journal} {Phys. Rev. C}\ }\textbf {\bibinfo {volume} {100}},\ \bibinfo
  {pages} {034325} (\bibinfo {year} {2019})}\BibitemShut {NoStop}%
\bibitem [{\citenamefont {Coraggio}\ \emph {et~al.}(2020)\citenamefont
  {Coraggio}, \citenamefont {Gargano}, \citenamefont {Itaco}, \citenamefont
  {Mancino},\ and\ \citenamefont {Nowacki}}]{Coraggio:2020}%
  \BibitemOpen
  \bibfield  {author} {\bibinfo {author} {\bibfnamefont {L.}~\bibnamefont
  {Coraggio}}, \bibinfo {author} {\bibfnamefont {A.}~\bibnamefont {Gargano}},
  \bibinfo {author} {\bibfnamefont {N.}~\bibnamefont {Itaco}}, \bibinfo
  {author} {\bibfnamefont {R.}~\bibnamefont {Mancino}},\ and\ \bibinfo {author}
  {\bibfnamefont {F.}~\bibnamefont {Nowacki}},\ }\bibfield  {title} {\bibinfo
  {title} {Calculation of the neutrinoless double-$\ensuremath{\beta}$ decay
  matrix element within the realistic shell model},\ }\href
  {https://doi.org/10.1103/PhysRevC.101.044315} {\bibfield  {journal} {\bibinfo
   {journal} {Phys. Rev. C}\ }\textbf {\bibinfo {volume} {101}},\ \bibinfo
  {pages} {044315} (\bibinfo {year} {2020})}\BibitemShut {NoStop}%
\bibitem [{\citenamefont {Deppisch}\ \emph {et~al.}(2020)\citenamefont
  {Deppisch}, \citenamefont {Graf}, \citenamefont {Iachello},\ and\
  \citenamefont {Kotila}}]{Deppisch:2020ztt}%
  \BibitemOpen
  \bibfield  {author} {\bibinfo {author} {\bibfnamefont {F.~F.}\ \bibnamefont
  {Deppisch}}, \bibinfo {author} {\bibfnamefont {L.}~\bibnamefont {Graf}},
  \bibinfo {author} {\bibfnamefont {F.}~\bibnamefont {Iachello}},\ and\
  \bibinfo {author} {\bibfnamefont {J.}~\bibnamefont {Kotila}},\ }\bibfield
  {title} {\bibinfo {title} {Analysis of light neutrino exchange and
  short-range mechanisms in
  $0\ensuremath{\nu}\ensuremath{\beta}\ensuremath{\beta}$ decay},\ }\href
  {https://doi.org/10.1103/PhysRevD.102.095016} {\bibfield  {journal} {\bibinfo
   {journal} {Phys. Rev. D}\ }\textbf {\bibinfo {volume} {102}},\ \bibinfo
  {pages} {095016} (\bibinfo {year} {2020})}\BibitemShut {NoStop}%
\bibitem [{\citenamefont {Engel}\ and\ \citenamefont
  {Men{\'e}ndez}(2017)}]{em}%
  \BibitemOpen
  \bibfield  {author} {\bibinfo {author} {\bibfnamefont {J.}~\bibnamefont
  {Engel}}\ and\ \bibinfo {author} {\bibfnamefont {J.}~\bibnamefont
  {Men{\'e}ndez}},\ }\bibfield  {title} {\bibinfo {title} {Status and future of
  nuclear matrix elements for neutrinoless double-beta decay: a review},\
  }\href@noop {} {\bibfield  {journal} {\bibinfo  {journal} {Rep. Prog. Phys.}\
  }\textbf {\bibinfo {volume} {80}},\ \bibinfo {pages} {046301} (\bibinfo
  {year} {2017})}\BibitemShut {NoStop}%
\bibitem [{\citenamefont {Yao}\ \emph {et~al.}(2020)\citenamefont {Yao},
  \citenamefont {Bally}, \citenamefont {Engel}, \citenamefont {Wirth},
  \citenamefont {Rodr{\'{i}}guez},\ and\ \citenamefont {Hergert}}]{Yao2020}%
  \BibitemOpen
  \bibfield  {author} {\bibinfo {author} {\bibfnamefont {J.~M.}\ \bibnamefont
  {Yao}}, \bibinfo {author} {\bibfnamefont {B.}~\bibnamefont {Bally}}, \bibinfo
  {author} {\bibfnamefont {J.}~\bibnamefont {Engel}}, \bibinfo {author}
  {\bibfnamefont {R.}~\bibnamefont {Wirth}}, \bibinfo {author} {\bibfnamefont
  {T.~R.}\ \bibnamefont {Rodr{\'{i}}guez}},\ and\ \bibinfo {author}
  {\bibfnamefont {H.}~\bibnamefont {Hergert}},\ }\bibfield  {title} {\bibinfo
  {title} {Ab initio treatment of collective correlations and the neutrinoless
  double beta decay of $^{48}\mathrm{Ca}$},\ }\href
  {https://doi.org/10.1103/PhysRevLett.124.232501} {\bibfield  {journal}
  {\bibinfo  {journal} {Phys. Rev. Lett.}\ }\textbf {\bibinfo {volume} {124}},\
  \bibinfo {pages} {232501} (\bibinfo {year} {2020})},\ \Eprint
  {https://arxiv.org/abs/1908.05424} {1908.05424} \BibitemShut {NoStop}%
\bibitem [{\citenamefont {Belley}\ \emph {et~al.}(2021)\citenamefont {Belley},
  \citenamefont {Payne}, \citenamefont {Stroberg}, \citenamefont {Miyagi},\
  and\ \citenamefont {Holt}}]{Belley2021}%
  \BibitemOpen
  \bibfield  {author} {\bibinfo {author} {\bibfnamefont {A.}~\bibnamefont
  {Belley}}, \bibinfo {author} {\bibfnamefont {C.~G.}\ \bibnamefont {Payne}},
  \bibinfo {author} {\bibfnamefont {S.~R.}\ \bibnamefont {Stroberg}}, \bibinfo
  {author} {\bibfnamefont {T.}~\bibnamefont {Miyagi}},\ and\ \bibinfo {author}
  {\bibfnamefont {J.~D.}\ \bibnamefont {Holt}},\ }\bibfield  {title} {\bibinfo
  {title} {{Ab Initio Neutrinoless Double-Beta Decay Matrix Elements for
  $^{48}$Ca , $^{76}$Ge, and $^{82}$Se}},\ }\href
  {https://doi.org/10.1103/PhysRevLett.126.042502} {\bibfield  {journal}
  {\bibinfo  {journal} {Phys. Rev. Lett.}\ }\textbf {\bibinfo {volume} {126}},\
  \bibinfo {pages} {042502} (\bibinfo {year} {2021})}\BibitemShut {NoStop}%
\bibitem [{\citenamefont {Novario}\ \emph {et~al.}(2020)\citenamefont
  {Novario}, \citenamefont {Gysbers}, \citenamefont {Engel}, \citenamefont
  {Hagen}, \citenamefont {Jansen}, \citenamefont {Morris}, \citenamefont
  {Navr\'atil}, \citenamefont {Papenbrock},\ and\ \citenamefont
  {Quaglioni}}]{Novario:2020dmr}%
  \BibitemOpen
  \bibfield  {author} {\bibinfo {author} {\bibfnamefont {S.}~\bibnamefont
  {Novario}}, \bibinfo {author} {\bibfnamefont {P.}~\bibnamefont {Gysbers}},
  \bibinfo {author} {\bibfnamefont {J.}~\bibnamefont {Engel}}, \bibinfo
  {author} {\bibfnamefont {G.}~\bibnamefont {Hagen}}, \bibinfo {author}
  {\bibfnamefont {G.~R.}\ \bibnamefont {Jansen}}, \bibinfo {author}
  {\bibfnamefont {T.~D.}\ \bibnamefont {Morris}}, \bibinfo {author}
  {\bibfnamefont {P.}~\bibnamefont {Navr\'atil}}, \bibinfo {author}
  {\bibfnamefont {T.}~\bibnamefont {Papenbrock}},\ and\ \bibinfo {author}
  {\bibfnamefont {S.}~\bibnamefont {Quaglioni}},\ }\bibfield  {title} {\bibinfo
  {title} {{Coupled-cluster calculations of neutrinoless double-beta decay in
  $^{48}$Ca}},\ }\Eprint {https://arxiv.org/abs/2008.09696} {arXiv:2008.09696
  [nucl-th]}  (\bibinfo {year} {2020})\BibitemShut {NoStop}%
\bibitem [{\citenamefont {Cirigliano}\ \emph {et~al.}(2018)\citenamefont
  {Cirigliano}, \citenamefont {Dekens}, \citenamefont {de~Vries}, \citenamefont
  {Graesser}, \citenamefont {Mereghetti}, \citenamefont {Pastore},\ and\
  \citenamefont {van Kolck}}]{Cirigliano:2018}%
  \BibitemOpen
  \bibfield  {author} {\bibinfo {author} {\bibfnamefont {V.}~\bibnamefont
  {Cirigliano}}, \bibinfo {author} {\bibfnamefont {W.}~\bibnamefont {Dekens}},
  \bibinfo {author} {\bibfnamefont {J.}~\bibnamefont {de~Vries}}, \bibinfo
  {author} {\bibfnamefont {M.~L.}\ \bibnamefont {Graesser}}, \bibinfo {author}
  {\bibfnamefont {E.}~\bibnamefont {Mereghetti}}, \bibinfo {author}
  {\bibfnamefont {S.}~\bibnamefont {Pastore}},\ and\ \bibinfo {author}
  {\bibfnamefont {U.}~\bibnamefont {van Kolck}},\ }\bibfield  {title} {\bibinfo
  {title} {New leading contribution to neutrinoless double-$\ensuremath{\beta}$
  decay},\ }\href {https://doi.org/10.1103/PhysRevLett.120.202001} {\bibfield
  {journal} {\bibinfo  {journal} {Phys. Rev. Lett.}\ }\textbf {\bibinfo
  {volume} {120}},\ \bibinfo {pages} {202001} (\bibinfo {year}
  {2018})}\BibitemShut {NoStop}%
\bibitem [{\citenamefont {Cirigliano}\ \emph {et~al.}(2019)\citenamefont
  {Cirigliano}, \citenamefont {Dekens}, \citenamefont {de~Vries}, \citenamefont
  {Graesser}, \citenamefont {Mereghetti}, \citenamefont {Pastore},
  \citenamefont {Piarulli}, \citenamefont {van Kolck},\ and\ \citenamefont
  {Wiringa}}]{Cirigliano2019}%
  \BibitemOpen
  \bibfield  {author} {\bibinfo {author} {\bibfnamefont {V.}~\bibnamefont
  {Cirigliano}}, \bibinfo {author} {\bibfnamefont {W.}~\bibnamefont {Dekens}},
  \bibinfo {author} {\bibfnamefont {J.}~\bibnamefont {de~Vries}}, \bibinfo
  {author} {\bibfnamefont {M.~L.}\ \bibnamefont {Graesser}}, \bibinfo {author}
  {\bibfnamefont {E.}~\bibnamefont {Mereghetti}}, \bibinfo {author}
  {\bibfnamefont {S.}~\bibnamefont {Pastore}}, \bibinfo {author} {\bibfnamefont
  {M.}~\bibnamefont {Piarulli}}, \bibinfo {author} {\bibfnamefont
  {U.}~\bibnamefont {van Kolck}},\ and\ \bibinfo {author} {\bibfnamefont
  {R.~B.}\ \bibnamefont {Wiringa}},\ }\bibfield  {title} {\bibinfo {title}
  {{Renormalized approach to neutrinoless double-$\beta$ decay}},\ }\href
  {https://doi.org/10.1103/PhysRevC.100.055504} {\bibfield  {journal} {\bibinfo
   {journal} {Phys. Rev. C}\ }\textbf {\bibinfo {volume} {100}},\ \bibinfo
  {pages} {055504} (\bibinfo {year} {2019})},\ \Eprint
  {https://arxiv.org/abs/1907.11254} {arXiv:1907.11254} \BibitemShut {NoStop}%
\bibitem [{\citenamefont {Cirigliano}\ \emph
  {et~al.}(2021{\natexlab{a}})\citenamefont {Cirigliano}, \citenamefont
  {Dekens}, \citenamefont {de~Vries}, \citenamefont {Hoferichter},\ and\
  \citenamefont {Mereghetti}}]{Cirigliano2020}%
  \BibitemOpen
  \bibfield  {author} {\bibinfo {author} {\bibfnamefont {V.}~\bibnamefont
  {Cirigliano}}, \bibinfo {author} {\bibfnamefont {W.}~\bibnamefont {Dekens}},
  \bibinfo {author} {\bibfnamefont {J.}~\bibnamefont {de~Vries}}, \bibinfo
  {author} {\bibfnamefont {M.}~\bibnamefont {Hoferichter}},\ and\ \bibinfo
  {author} {\bibfnamefont {E.}~\bibnamefont {Mereghetti}},\ }\bibfield  {title}
  {\bibinfo {title} {{Toward Complete Leading-Order Predictions for
  Neutrinoless Double $\beta$ Decay}},\ }\href
  {https://doi.org/10.1103/PhysRevLett.126.172002} {\bibfield  {journal}
  {\bibinfo  {journal} {Phys. Rev. Lett.}\ }\textbf {\bibinfo {volume} {126}},\
  \bibinfo {pages} {172002} (\bibinfo {year} {2021}{\natexlab{a}})},\ \Eprint
  {https://arxiv.org/abs/2012.11602} {arXiv:2012.11602} \BibitemShut {NoStop}%
\bibitem [{\citenamefont {Cirigliano}\ \emph
  {et~al.}(2021{\natexlab{b}})\citenamefont {Cirigliano}, \citenamefont
  {Dekens}, \citenamefont {de~Vries}, \citenamefont {Hoferichter},\ and\
  \citenamefont {Mereghetti}}]{Cirigliano2021}%
  \BibitemOpen
  \bibfield  {author} {\bibinfo {author} {\bibfnamefont {V.}~\bibnamefont
  {Cirigliano}}, \bibinfo {author} {\bibfnamefont {W.}~\bibnamefont {Dekens}},
  \bibinfo {author} {\bibfnamefont {J.}~\bibnamefont {de~Vries}}, \bibinfo
  {author} {\bibfnamefont {M.}~\bibnamefont {Hoferichter}},\ and\ \bibinfo
  {author} {\bibfnamefont {E.}~\bibnamefont {Mereghetti}},\ }\bibfield  {title}
  {\bibinfo {title} {{Determining the leading-order contact term in
  neutrinoless double $\beta$ decay}},\ }\Eprint
  {https://arxiv.org/abs/2102.03371} {arXiv:2102.03371}  (\bibinfo {year}
  {2021}{\natexlab{b}})\BibitemShut {NoStop}%
\bibitem [{\citenamefont {Cottingham}(1963)}]{Cottingham1963}%
  \BibitemOpen
  \bibfield  {author} {\bibinfo {author} {\bibfnamefont {W.~N.}\ \bibnamefont
  {Cottingham}},\ }\bibfield  {title} {\bibinfo {title} {{The neutron proton
  mass difference and electron scattering experiments}},\ }\href
  {https://doi.org/10.1016/0003-4916(63)90023-X} {\bibfield  {journal}
  {\bibinfo  {journal} {Ann. Phys. (N. Y).}\ }\textbf {\bibinfo {volume}
  {25}},\ \bibinfo {pages} {424} (\bibinfo {year} {1963})}\BibitemShut
  {NoStop}%
\bibitem [{\citenamefont {\ifmmode~\check{S}\else \v{S}\fi{}imkovic}\ \emph
  {et~al.}(1999)\citenamefont {\ifmmode~\check{S}\else \v{S}\fi{}imkovic},
  \citenamefont {Pantis}, \citenamefont {Vergados},\ and\ \citenamefont
  {Faessler}}]{Simkovic:1999}%
  \BibitemOpen
  \bibfield  {author} {\bibinfo {author} {\bibfnamefont {F.}~\bibnamefont
  {\ifmmode~\check{S}\else \v{S}\fi{}imkovic}}, \bibinfo {author}
  {\bibfnamefont {G.}~\bibnamefont {Pantis}}, \bibinfo {author} {\bibfnamefont
  {J.~D.}\ \bibnamefont {Vergados}},\ and\ \bibinfo {author} {\bibfnamefont
  {A.}~\bibnamefont {Faessler}},\ }\bibfield  {title} {\bibinfo {title}
  {Additional nucleon current contributions to neutrinoless double
  $\ensuremath{\beta}$ decay},\ }\href
  {https://doi.org/10.1103/PhysRevC.60.055502} {\bibfield  {journal} {\bibinfo
  {journal} {Phys. Rev. C}\ }\textbf {\bibinfo {volume} {60}},\ \bibinfo
  {pages} {055502} (\bibinfo {year} {1999})}\BibitemShut {NoStop}%
\bibitem [{\citenamefont {Descouvemont}\ and\ \citenamefont
  {Baye}(2010)}]{Descouvemont2010}%
  \BibitemOpen
  \bibfield  {author} {\bibinfo {author} {\bibfnamefont {P.}~\bibnamefont
  {Descouvemont}}\ and\ \bibinfo {author} {\bibfnamefont {D.}~\bibnamefont
  {Baye}},\ }\bibfield  {title} {\bibinfo {title} {{The R-matrix theory}},\
  }\href {https://doi.org/10.1088/0034-4885/73/3/036301} {\bibfield  {journal}
  {\bibinfo  {journal} {Rep. Prog. Phys.}\ }\textbf {\bibinfo {volume} {73}},\
  \bibinfo {pages} {036301} (\bibinfo {year} {2010})}\BibitemShut {NoStop}%
\bibitem [{Sup()}]{SupplementalMaterial}%
  \BibitemOpen
  \href@noop {} {}\bibinfo {note} {{See Supplemental Material at [URL will be
  inserted by publisher] for additional validation details, recommended LEC
  values for multiple interactions at various SRG scales, and NME
  figures.}}\BibitemShut {Stop}%
\bibitem [{\citenamefont {Entem}\ and\ \citenamefont
  {Machleidt}(2003)}]{Entem2003}%
  \BibitemOpen
  \bibfield  {author} {\bibinfo {author} {\bibfnamefont {D.~R.}\ \bibnamefont
  {Entem}}\ and\ \bibinfo {author} {\bibfnamefont {R.}~\bibnamefont
  {Machleidt}},\ }\bibfield  {title} {\bibinfo {title} {{Accurate
  charge-dependent nucleon-nucleon potential at fourth order of chiral
  perturbation theory}},\ }\href {https://doi.org/10.1103/PhysRevC.68.041001}
  {\bibfield  {journal} {\bibinfo  {journal} {Phys. Rev. C}\ }\textbf {\bibinfo
  {volume} {68}},\ \bibinfo {pages} {041001(R)} (\bibinfo {year} {2003})},\
  \Eprint {https://arxiv.org/abs/0304018} {arXiv:0304018 [nucl-th]}
  \BibitemShut {NoStop}%
\bibitem [{\citenamefont {Entem}\ \emph {et~al.}(2017)\citenamefont {Entem},
  \citenamefont {Machleidt},\ and\ \citenamefont {Nosyk}}]{Entem2017}%
  \BibitemOpen
  \bibfield  {author} {\bibinfo {author} {\bibfnamefont {D.~R.}\ \bibnamefont
  {Entem}}, \bibinfo {author} {\bibfnamefont {R.}~\bibnamefont {Machleidt}},\
  and\ \bibinfo {author} {\bibfnamefont {Y.}~\bibnamefont {Nosyk}},\ }\bibfield
   {title} {\bibinfo {title} {{High-quality two-nucleon potentials up to fifth
  order of the chiral expansion}},\ }\href
  {https://doi.org/10.1103/PhysRevC.96.024004} {\bibfield  {journal} {\bibinfo
  {journal} {Phys. Rev. C}\ }\textbf {\bibinfo {volume} {96}},\ \bibinfo
  {pages} {024004} (\bibinfo {year} {2017})},\ \Eprint
  {https://arxiv.org/abs/1703.05454} {arXiv:1703.05454} \BibitemShut {NoStop}%
\bibitem [{\citenamefont {Jiang}\ \emph {et~al.}(2020)\citenamefont {Jiang},
  \citenamefont {Ekstr{\"{o}}m}, \citenamefont {Forss{\'{e}}n}, \citenamefont
  {Hagen}, \citenamefont {Jansen},\ and\ \citenamefont
  {Papenbrock}}]{Jiang2020}%
  \BibitemOpen
  \bibfield  {author} {\bibinfo {author} {\bibfnamefont {W.~G.}\ \bibnamefont
  {Jiang}}, \bibinfo {author} {\bibfnamefont {A.}~\bibnamefont
  {Ekstr{\"{o}}m}}, \bibinfo {author} {\bibfnamefont {C.}~\bibnamefont
  {Forss{\'{e}}n}}, \bibinfo {author} {\bibfnamefont {G.}~\bibnamefont
  {Hagen}}, \bibinfo {author} {\bibfnamefont {G.~R.}\ \bibnamefont {Jansen}},\
  and\ \bibinfo {author} {\bibfnamefont {T.}~\bibnamefont {Papenbrock}},\
  }\bibfield  {title} {\bibinfo {title} {{Accurate bulk properties of nuclei
  from $A=2$ to $\infty$ from potentials with $\Delta$ isobars}},\ }\href
  {https://doi.org/10.1103/PhysRevC.102.054301} {\bibfield  {journal} {\bibinfo
   {journal} {Phys. Rev. C}\ }\textbf {\bibinfo {volume} {102}},\ \bibinfo
  {pages} {054301} (\bibinfo {year} {2020})},\ \Eprint
  {https://arxiv.org/abs/2006.16774} {arXiv:2006.16774} \BibitemShut {NoStop}%
\bibitem [{\citenamefont {Yao}\ \emph {et~al.}(2021)\citenamefont {Yao},
  \citenamefont {Belley}, \citenamefont {Wirth}, \citenamefont {Miyagi},
  \citenamefont {Payne}, \citenamefont {Stroberg}, \citenamefont {Hergert},\
  and\ \citenamefont {Holt}}]{Yao2021}%
  \BibitemOpen
  \bibfield  {author} {\bibinfo {author} {\bibfnamefont {J.~M.}\ \bibnamefont
  {Yao}}, \bibinfo {author} {\bibfnamefont {A.}~\bibnamefont {Belley}},
  \bibinfo {author} {\bibfnamefont {R.}~\bibnamefont {Wirth}}, \bibinfo
  {author} {\bibfnamefont {T.}~\bibnamefont {Miyagi}}, \bibinfo {author}
  {\bibfnamefont {C.~G.}\ \bibnamefont {Payne}}, \bibinfo {author}
  {\bibfnamefont {S.~R.}\ \bibnamefont {Stroberg}}, \bibinfo {author}
  {\bibfnamefont {H.}~\bibnamefont {Hergert}},\ and\ \bibinfo {author}
  {\bibfnamefont {J.~D.}\ \bibnamefont {Holt}},\ }\bibfield  {title} {\bibinfo
  {title} {Ab initio benchmarks of neutrinoless double-$\beta$ decay in light
  nuclei with a chiral hamiltonian},\ }\href@noop {} {\bibfield  {journal}
  {\bibinfo  {journal} {Phys. Rev. C}\ }\textbf {\bibinfo {volume} {103}}
  (\bibinfo {year} {2021})}\BibitemShut {NoStop}%
\bibitem [{\citenamefont {G{\l}azek}\ and\ \citenamefont
  {Wilson}(1993)}]{Gazek1993}%
  \BibitemOpen
  \bibfield  {author} {\bibinfo {author} {\bibfnamefont {S.}~\bibnamefont
  {G{\l}azek}}\ and\ \bibinfo {author} {\bibfnamefont {K.}~\bibnamefont
  {Wilson}},\ }\bibfield  {title} {\bibinfo {title} {{Renormalization of
  Hamiltonians}},\ }\href {https://doi.org/10.1103/PhysRevD.48.5863} {\bibfield
   {journal} {\bibinfo  {journal} {Phys. Rev. D}\ }\textbf {\bibinfo {volume}
  {48}},\ \bibinfo {pages} {5863} (\bibinfo {year} {1993})}\BibitemShut
  {NoStop}%
\bibitem [{\citenamefont {Wegner}(1994)}]{Wegner1994}%
  \BibitemOpen
  \bibfield  {author} {\bibinfo {author} {\bibfnamefont {F.~J.}\ \bibnamefont
  {Wegner}},\ }\bibfield  {title} {\bibinfo {title} {{Flow-equations for
  Hamiltonians}},\ }\href {https://doi.org/10.1002/andp.19945060203} {\bibfield
   {journal} {\bibinfo  {journal} {Ann. Phys. (Leipzig)}\ }\textbf {\bibinfo
  {volume} {506}},\ \bibinfo {pages} {77} (\bibinfo {year} {1994})}\BibitemShut
  {NoStop}%
\bibitem [{\citenamefont {Bogner}\ \emph {et~al.}(2007)\citenamefont {Bogner},
  \citenamefont {Furnstahl},\ and\ \citenamefont {Perry}}]{Bogner2007}%
  \BibitemOpen
  \bibfield  {author} {\bibinfo {author} {\bibfnamefont {S.}~\bibnamefont
  {Bogner}}, \bibinfo {author} {\bibfnamefont {R.~J.}\ \bibnamefont
  {Furnstahl}},\ and\ \bibinfo {author} {\bibfnamefont {R.}~\bibnamefont
  {Perry}},\ }\bibfield  {title} {\bibinfo {title} {{Similarity renormalization
  group for nucleon-nucleon interactions}},\ }\href
  {https://doi.org/10.1103/PhysRevC.75.061001} {\bibfield  {journal} {\bibinfo
  {journal} {Phys. Rev. C}\ }\textbf {\bibinfo {volume} {75}},\ \bibinfo
  {pages} {061001} (\bibinfo {year} {2007})},\ \Eprint
  {https://arxiv.org/abs/0611045} {arXiv:0611045 [nucl-th]} \BibitemShut
  {NoStop}%
\bibitem [{\citenamefont {Hebeler}\ \emph {et~al.}(2011)\citenamefont
  {Hebeler}, \citenamefont {Bogner}, \citenamefont {Furnstahl}, \citenamefont
  {Nogga},\ and\ \citenamefont {Schwenk}}]{Hebeler2011}%
  \BibitemOpen
  \bibfield  {author} {\bibinfo {author} {\bibfnamefont {K.}~\bibnamefont
  {Hebeler}}, \bibinfo {author} {\bibfnamefont {S.~K.}\ \bibnamefont {Bogner}},
  \bibinfo {author} {\bibfnamefont {R.~J.}\ \bibnamefont {Furnstahl}}, \bibinfo
  {author} {\bibfnamefont {A.}~\bibnamefont {Nogga}},\ and\ \bibinfo {author}
  {\bibfnamefont {A.}~\bibnamefont {Schwenk}},\ }\bibfield  {title} {\bibinfo
  {title} {{Improved nuclear matter calculations from chiral low-momentum
  interactions}},\ }\href {https://doi.org/10.1103/PhysRevC.83.031301}
  {\bibfield  {journal} {\bibinfo  {journal} {Phys. Rev. C}\ }\textbf {\bibinfo
  {volume} {83}},\ \bibinfo {pages} {031301(R)} (\bibinfo {year}
  {2011})}\BibitemShut {NoStop}%
\bibitem [{\citenamefont {Som{\`{a}}}\ \emph {et~al.}(2020)\citenamefont
  {Som{\`{a}}}, \citenamefont {Navr{\'{a}}til}, \citenamefont {Raimondi},
  \citenamefont {Barbieri},\ and\ \citenamefont {Duguet}}]{Soma2020}%
  \BibitemOpen
  \bibfield  {author} {\bibinfo {author} {\bibfnamefont {V.}~\bibnamefont
  {Som{\`{a}}}}, \bibinfo {author} {\bibfnamefont {P.}~\bibnamefont
  {Navr{\'{a}}til}}, \bibinfo {author} {\bibfnamefont {F.}~\bibnamefont
  {Raimondi}}, \bibinfo {author} {\bibfnamefont {C.}~\bibnamefont {Barbieri}},\
  and\ \bibinfo {author} {\bibfnamefont {T.}~\bibnamefont {Duguet}},\
  }\bibfield  {title} {\bibinfo {title} {{Novel chiral Hamiltonian and
  observables in light and medium-mass nuclei}},\ }\href
  {https://doi.org/10.1103/PhysRevC.101.014318} {\bibfield  {journal} {\bibinfo
   {journal} {Phys. Rev. C}\ }\textbf {\bibinfo {volume} {101}},\ \bibinfo
  {pages} {014318} (\bibinfo {year} {2020})},\ \Eprint
  {https://arxiv.org/abs/1907.09790} {arXiv:1907.09790} \BibitemShut {NoStop}%
\bibitem [{\citenamefont {H{\"{u}}ther}\ \emph {et~al.}(2020)\citenamefont
  {H{\"{u}}ther}, \citenamefont {Vobig}, \citenamefont {Hebeler}, \citenamefont
  {Machleidt},\ and\ \citenamefont {Roth}}]{Huther2020}%
  \BibitemOpen
  \bibfield  {author} {\bibinfo {author} {\bibfnamefont {T.}~\bibnamefont
  {H{\"{u}}ther}}, \bibinfo {author} {\bibfnamefont {K.}~\bibnamefont {Vobig}},
  \bibinfo {author} {\bibfnamefont {K.}~\bibnamefont {Hebeler}}, \bibinfo
  {author} {\bibfnamefont {R.}~\bibnamefont {Machleidt}},\ and\ \bibinfo
  {author} {\bibfnamefont {R.}~\bibnamefont {Roth}},\ }\bibfield  {title}
  {\bibinfo {title} {{Family of chiral two- plus three-nucleon interactions for
  accurate nuclear structure studies}},\ }\href
  {https://doi.org/10.1016/j.physletb.2020.135651} {\bibfield  {journal}
  {\bibinfo  {journal} {Phys. Lett. B}\ }\textbf {\bibinfo {volume} {808}},\
  \bibinfo {pages} {135651} (\bibinfo {year} {2020})},\ \Eprint
  {https://arxiv.org/abs/1911.04955} {arXiv:1911.04955} \BibitemShut {NoStop}%
\bibitem [{\citenamefont {Roth}(2009)}]{Roth2009}%
  \BibitemOpen
  \bibfield  {author} {\bibinfo {author} {\bibfnamefont {R.}~\bibnamefont
  {Roth}},\ }\bibfield  {title} {\bibinfo {title} {{Importance truncation for
  large-scale configuration interaction approaches}},\ }\href
  {https://doi.org/10.1103/PhysRevC.79.064324} {\bibfield  {journal} {\bibinfo
  {journal} {Phys. Rev. C}\ }\textbf {\bibinfo {volume} {79}},\ \bibinfo
  {pages} {064324} (\bibinfo {year} {2009})},\ \Eprint
  {https://arxiv.org/abs/0903.4605} {arXiv:0903.4605} \BibitemShut {NoStop}%
\bibitem [{\citenamefont {Davoudi}\ and\ \citenamefont
  {Kadam}(2021)}]{Davoudi:2021}%
  \BibitemOpen
  \bibfield  {author} {\bibinfo {author} {\bibfnamefont {Z.}~\bibnamefont
  {Davoudi}}\ and\ \bibinfo {author} {\bibfnamefont {S.~V.}\ \bibnamefont
  {Kadam}},\ }\bibfield  {title} {\bibinfo {title} {{Path from Lattice QCD to
  the Short-Distance Contribution to
  $0\ensuremath{\nu}\ensuremath{\beta}\ensuremath{\beta}$ Decay with a Light
  Majorana Neutrino}},\ }\href {https://doi.org/10.1103/PhysRevLett.126.152003}
  {\bibfield  {journal} {\bibinfo  {journal} {Phys. Rev. Lett.}\ }\textbf
  {\bibinfo {volume} {126}},\ \bibinfo {pages} {152003} (\bibinfo {year}
  {2021})}\BibitemShut {NoStop}%
\end{thebibliography}%

\pagebreak

\section*{Supplemental Material}

\paragraph{Computational details.}
Since the long-range neutrino potential exhibits $1/r$ behavior and contributes at distances $r\gg a$, we split the amplitude integral at the $R$-matrix channel radius.
In the exterior region we use the asymptotic form of the wavefunction.
This approximation is exact in the $^1S_0$ channel without Coulomb interaction because the free solutions of the radial Schrödinger equation are proportional to sine and cosine functions.
For channels with higher angular momentum or ones that include the Coulomb interaction the difference between the full and asymptotic forms is negligible for our choice of the channel radius $a = \SI{15}{\fm}$.

\begin{figure}
  \centering
  \includegraphics{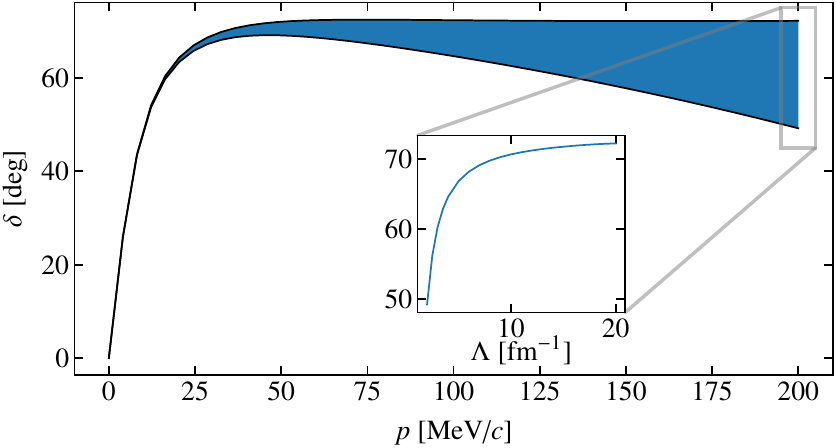}
  \caption{\label{fig:lo phaseshifts}%
    Leading-order NN phase shifts at different regulator cutoffs $\Lambda=\{ \SI{2}{\per\fm},\dotsc, \SI{20}{\per\fm} \}$ with $n_\text{exp}=2$.
    The interaction is adjusted to the neutron-proton scattering length.
  }
\end{figure}

\begin{figure}
  \centering
  \includegraphics{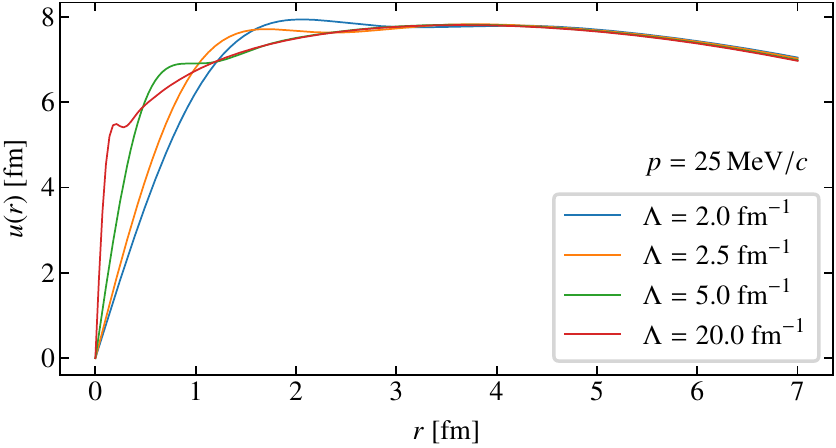}
  \caption{\label{fig:lo wavefunction}%
    Leading-order wavefunctions at different regulator cutoffs $\Lambda$ at momentum $p=\SI{25}{\MeVc}$ with $n_\text{exp}=2$.
  }
\end{figure}

\begin{figure}
  \centering
  \includegraphics{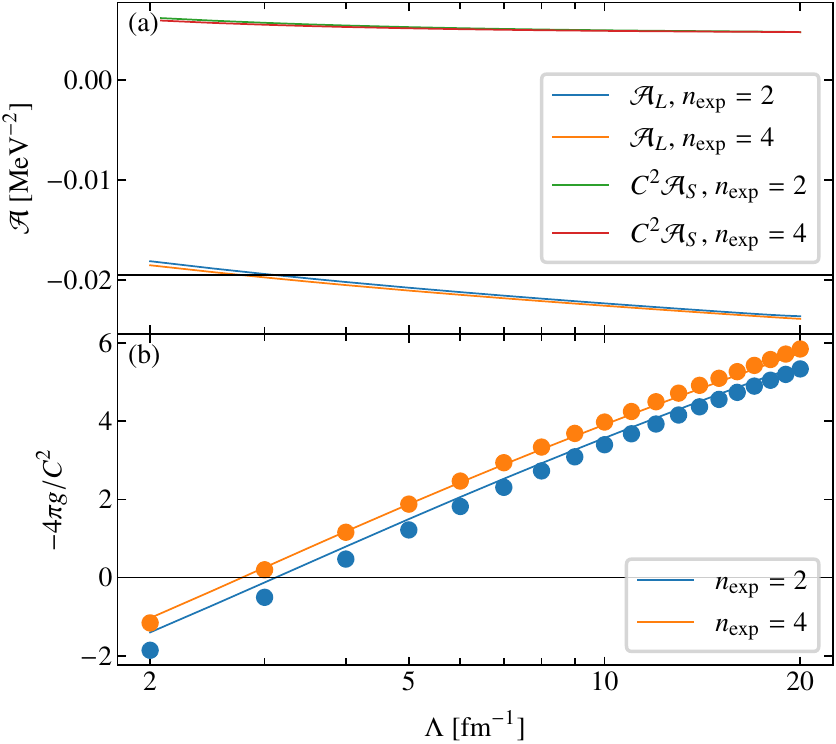}
  \caption{\label{fig:amplitude lambda dependence}%
    (a) Dependence of the short- and long-range parts of the amplitude on the regulator scale $\Lambda$ at the kinematic point $p=\SI{25}{\MeVc}, p'=\SI{30}{\MeVc}$.
    (b) Dependence of the short-range LEC on the regulator scale $\Lambda$ at the kinematic point $p=\SI{25}{\MeVc}, p'=\SI{30}{\MeVc}$.
    Lines show our results, the dots mark results taken from Ref.\ \cite{Cirigliano2021}.
}
\end{figure}

\paragraph{Validation.}
We use a leading-order isospin-symmetric potential to verify our calculation against the results shown in \cite{Cirigliano2021}.
In the $^1S_0$ channel, the momentum-space form of this potential reads
\begin{align}
  V_{LO}(p,p') &= \frac{1}{2\pi^2} \frac{g_A^2}{4f_\pi^2} \biggl\{C  -  \frac{m_\pi^2}{4 p p'} \log\biggl[1 + \frac{4 p p'}{m_\pi^2 + (p-p')^2}\biggr]\biggr\}
\notag\\&\hphantom{{}={}}\times
  \exp\biggl[-\biggl(\frac{p}{\Lambda}\biggr)^{2n_\text{exp}}\biggr] \exp\biggl[-\biggl(\frac{p'}{\Lambda}\biggr)^{2n_\text{exp}}\biggr].
\end{align}
The dimensionless LEC $C$ is adjusted to reproduce the neu\-tron--proton scattering length $a_{np}=\SI{-23.74}{\fm}$.

To validate our calculations, we generate a set of interactions with cutoffs $\Lambda$ from $\SI{2}{\per\fm}$ to $\SI{20}{\per\fm}$, setting the regulator exponent to $n_\text{exp}=2$ and $4$.
The phase shifts up to a relative momentum of \SI{200}{\MeVc} are shown in \cref{fig:lo phaseshifts}. It is shown that the phase shifts are only weakly dependent on the regulator cutoffs in the low-momentum region with $p<\SI{50}{\MeVc}$, consistent with the findings in Ref.~\cite{Cirigliano2019}.

\Cref{fig:lo wavefunction} shows the nucleon-nucleon wavefunction for different regulator cutoffs.
At short ranges the wavefunctions exhibit a clear cutoff dependence, rising more quickly for higher cutoffs.
Beyond a relative distance of \SI{3}{\fm} they all collapse to the same curve, because the phase shift at this momentum is approximately cutoff independent.

With the scattering wavefunctions calculated from this set of interactions, we compute the long- and short-range amplitudes at the kinematic point $p=\SI{25}{\MeVc}, p'=\SI{30}{\MeVc}$ \cite{Cirigliano2020,Cirigliano2021}, shown in \cref{fig:amplitude lambda dependence}(a).
The long-range part shows a logarithmic dependence on the regulator cutoff $\Lambda$, while the combination $C^2 \mathcal{A}_S$, computed using the same regulator parameters as the interaction, is virtually independent of it.

By requiring that the total amplitude matches the synthetic datum, this implies that the ratio of LECs $g/C^2$ exhibits the same logarithmic scale dependence.
The dependence is shown in \cref{fig:amplitude lambda dependence}(b).
There is a small discrepancy between our LECs and the ones taken from Ref.~\cite{Cirigliano2021} for $n_\text{exp}=2$, which might be attributable to the different solution methods for the scattering problem or the choice of momentum- or coordinate-space grids.
The discrepancy becomes smaller with increasing regulator power $n_\text{exp}$, in particular for the values $n_\text{exp}=\{3,4\}$ used in the remainder of this work.

\begin{figure}
  \centering
  \includegraphics{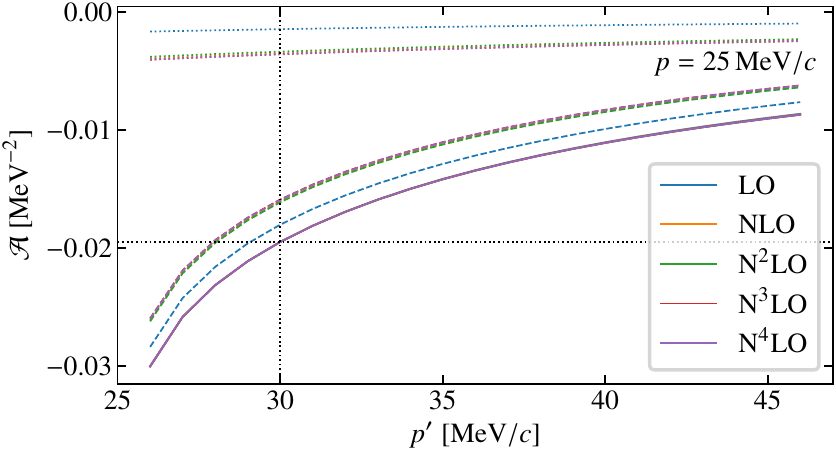}
  \caption{\label{fig:order dependence momentum}%
    Momentum dependence of the short- and long-range parts, as well as the total amplitude for the EMN potential at different orders in the chiral EFT expansion.
    Shown are the scaled short-range part $-2g\mathcal{A}_{S}$ (dotted lines), the long-range part $\mathcal{A}_{L}$ (dashed lines), and the total amplitude $\mathcal{A}_{L} - 2g\mathcal{A}_{S}$ (solid lines).
    The dotted black lines mark the synthetic datum.
    The variation with respect to order over the momentum range shown is less than \SI{1}{\percent} and rapidly converging with increasing order.
  }
\end{figure}

\paragraph{Order-by-order momentum dependence.}
\Cref{fig:order dependence momentum} shows the amplitudes for the different chiral orders.
The long-range part of the amplitude converges quickly beyond LO: the variation between orders over the momentum range shown is less than \SI{3}{\percent}, and it decreases when going to higher orders.
The short-range part shows a much larger variation of \SI{40}{\percent}, but that dependence is again an order-dependent scaling factor that can be absorbed into the LEC. Thus, the total amplitude is essentially converged at leading order for low momenta.

The long-range part of the LO interaction is larger in magnitude than the higher-order results.
This difference is because the LO interaction breaks charge symmetry with significantly different $nn$ and $pp$ scattering lengths (even without the Coulomb interaction).

\begin{figure}
  \centering
  \includegraphics[width=0.85\columnwidth]{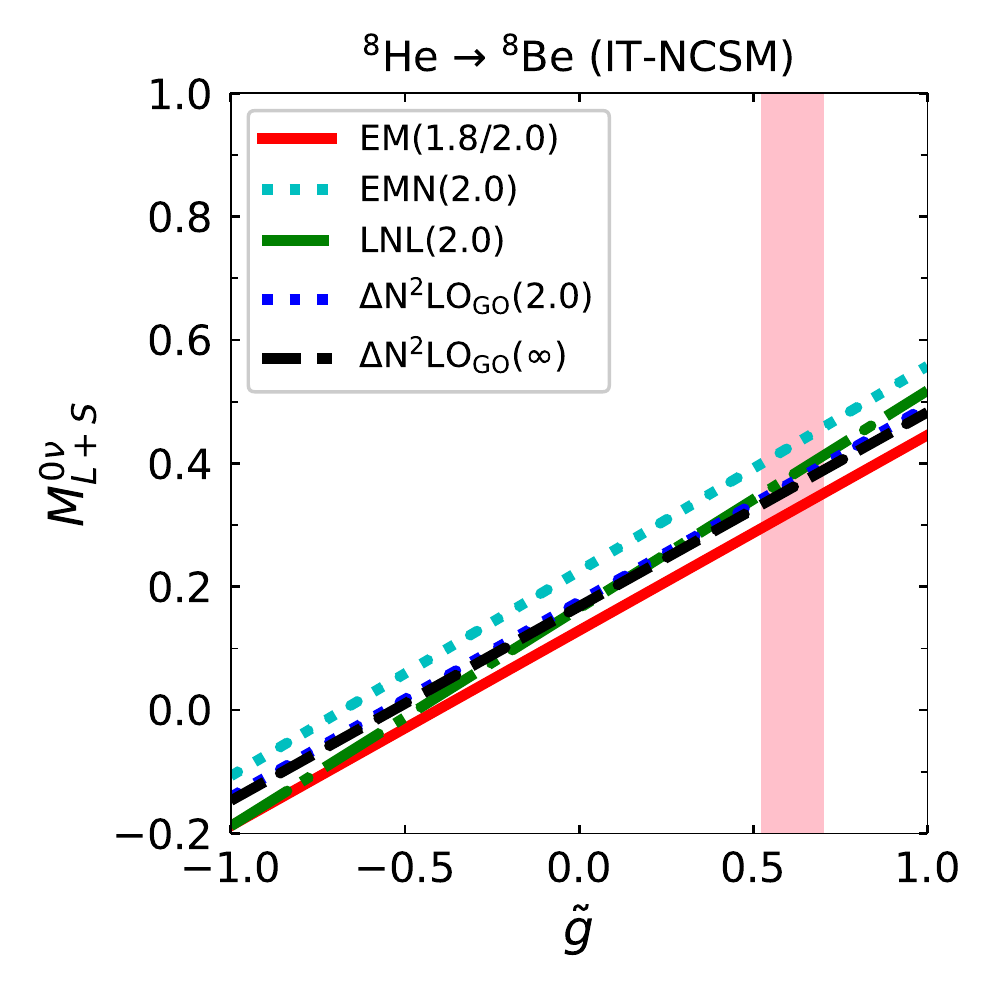}
  \caption{\label{fig:NME4A8}%
   The NMEs of neutrinoless double beta decay for  \nuclide[8]{He} from the IT-NCSM calculation using different chiral interactions.
  }
\end{figure}

\begin{figure}
  \centering
  \includegraphics[width=0.85\columnwidth]{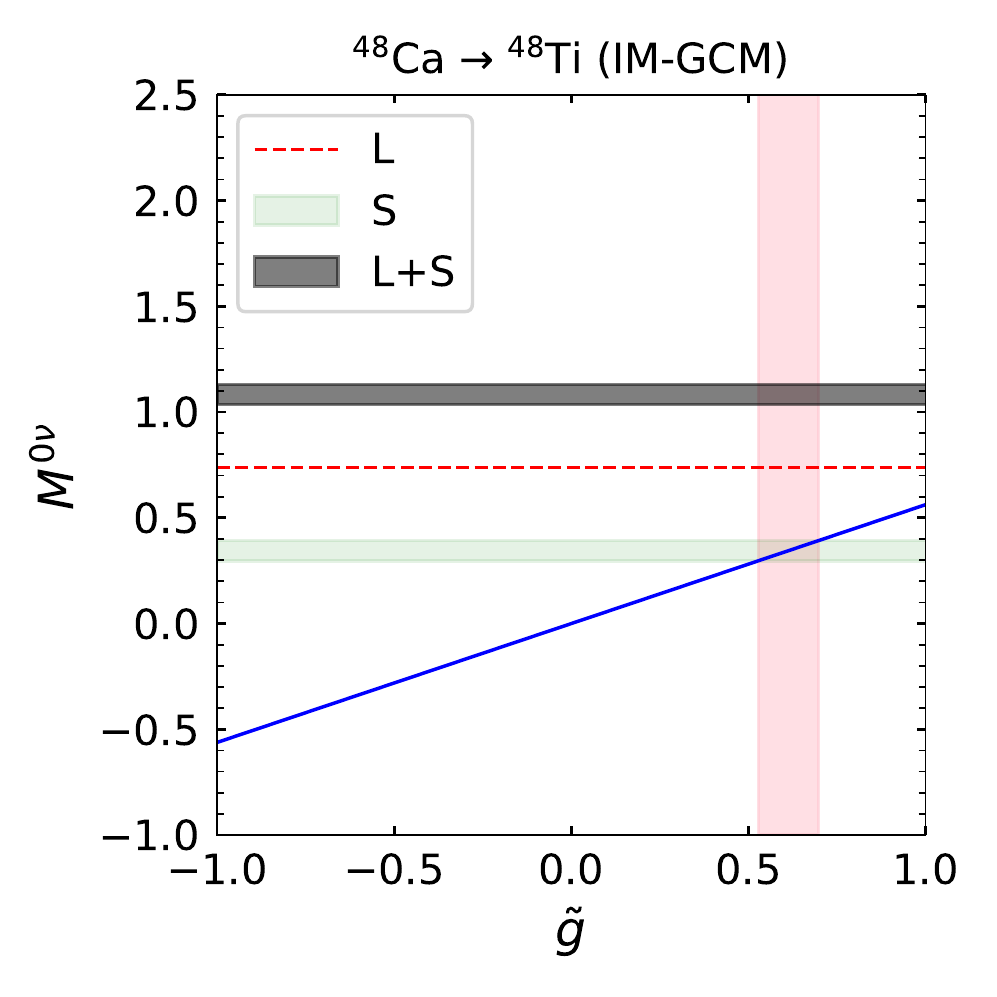}
  \caption{\label{fig:NME4Ca}%
   The NMEs of neutrinoless double beta decay for  \nuclide[48]{Ca} from the IM-GCM calculation using $e_{\text{Max}}=8$, and $\hbar\omega=\SI{16}{\MeV}$. The blue line indicates the contribution of the short-range contact operator to the NME, $4\pi R/(g_A^2) \braket{\nuclide[48]{Ti}| (- 2\tilde g)\hat{V}_{S}|\nuclide[48]{Ca}}$.
  }
\end{figure}

\paragraph{Amplitude tables.}
\Cref{tab:ENM LECs} and \cref{tab:N2LOGO LECs} show a compilation of the long- and short-range amplitudes at the kinematic point $p=\SI{25}{\MeVc}$, $p'=\SI{30}{\MeVc}$ for the EMN and $\Delta$N\textsuperscript2LO$_\text{GO}$ interactions.
The amplitudes are computed at different chiral orders (for the EMN) and various SRG scales.
The LECs $g,\tilde{g}$ are obtained by imposing the synthetic datum $\mathcal{A}=\SI{-0.195(5)}{\MeV\tothe{-2}}$.
The LEC uncertainty $\Delta g$ reflects the uncertainty in the synthetic datum and is the same for $g$ and $\tilde{g}$.

\paragraph{NMEs for finite nuclei.}
\Cref{fig:NME4A8} shows the dependence of the $\nuclide[8]{He}\to\nuclide[8]{Be}$ NME on the contact LEC $\tilde{g}$ from the IT-NCSM calculation.
The value of the contact is similar for the different interactions employed and indicated by the vertical band.
The contact term is slightly larger for the LNL Hamiltonian than for the others, leading to a steeper dependence on $\tilde{g}$.
The total NMEs $M^{0\nu}_{L+S}$ are very similar, except that the curve for the EMN(2.0) is shifted up due to a larger long-range part.
A similar plot is shown in \cref{fig:NME4Ca}, showing the \nuclide[48]{Ca} NME for the EM1.8/2.0 Hamiltonian with $e_{\rm Max}=8$, and $\hbar\omega=\SI{16}{\MeV}$ from the IM-GCM calculation \cite{Yao2020}.
Here, the long- and short-range parts are separated.
The blue line shows the LEC dependence of the short-range part, whose intersection with the vertical band, showing the uncertainty in $\tilde{g}$, yields an uncertainty band (green) for the contact contribution.
The dashed line marks the contribution of the long-range part, and the gray band shows their sum.
The detailed values of the NME for $\nuclide[6]{He}\to\nuclide[6]{Be}$, $\nuclide[8]{He}\to\nuclide[8]{Be}$ and $\nuclide[48]{Ca}\to\nuclide[48]{Ti}$ are given in Table \ref{tab:nme4all}. It is shown that the NMEs by different interactions overlap with each other.

\begin{table}[H]
    \centering
        \caption{\label{tab:ENM LECs}%
    Amplitudes and LECs for the EMN family of interactions at different orders and SRG scales.
    Values for the SRG scale are given in units of \si{\per\fm}, amplitudes in units of \si{\MeV\tothe{-2}}.
    The contact term is regularized using $\Lambda=\SI{500}{\MeVc}$ and $n_\text{exp}=3$.
    The quantities with a tilde incorporate beyond-LO effects in the operator.
    }
    \begin{tabular}{lccccccc}
\toprule
Order & $\lambda$ & $10^3 \tilde{\mathcal{A}}_L$ & $10^3 \mathcal{A}_L$ & $10^3 \mathcal{A}_S$ & $\tilde{g}$ & $g$ & $\Delta g$ \\
\midrule
LO & $\infty$ & $-17.653$ & $-18.034$ & $8.4847$ & $0.109$ & $0.086$ & $0.029$ \\
   & $2.50$   & $-17.440$ & $-17.752$ & $7.6083$ & $0.135$ & $0.115$ & $0.033$ \\
   & $2.24$   & $-17.353$ & $-17.643$ & $7.2423$ & $0.148$ & $0.128$ & $0.035$ \\
   & $2.20$   & $-17.338$ & $-17.625$ & $7.1811$ & $0.151$ & $0.131$ & $0.035$ \\
   & $2.00$   & $-17.241$ & $-17.508$ & $6.7840$ & $0.166$ & $0.147$ & $0.037$ \\
   & $1.88$   & $-17.168$ & $-17.421$ & $6.4950$ & $0.180$ & $0.160$ & $0.038$ \\
   & $1.80$   & $-17.111$ & $-17.354$ & $6.2779$ & $0.190$ & $0.171$ & $0.040$ \\
\midrule
NLO & $\infty$ & $-15.951$ & $-16.020$ & $2.6403$ & $0.672$ & $0.659$ & $0.095$ \\
    & $2.50$   & $-16.005$ & $-16.111$ & $2.9479$ & $0.593$ & $0.575$ & $0.085$ \\
    & $2.24$   & $-16.006$ & $-16.117$ & $2.9982$ & $0.583$ & $0.564$ & $0.083$ \\
    & $2.20$   & $-16.005$ & $-16.117$ & $3.0032$ & $0.582$ & $0.563$ & $0.083$ \\
    & $2.00$   & $-15.994$ & $-16.108$ & $3.0146$ & $0.582$ & $0.563$ & $0.083$ \\
    & $1.88$   & $-15.980$ & $-16.094$ & $3.0033$ & $0.586$ & $0.567$ & $0.083$ \\
    & $1.80$   & $-15.966$ & $-16.080$ & $2.9855$ & $0.592$ & $0.573$ & $0.084$ \\
\midrule
N\textsuperscript2LO & $\infty$ & $-16.020$ & $-16.131$ & $3.3533$ & $0.519$ & $0.502$ & $0.075$ \\
                     & $2.50$   & $-16.007$ & $-16.132$ & $3.3458$ & $0.522$ & $0.503$ & $0.075$ \\
                     & $2.24$   & $-15.991$ & $-16.117$ & $3.3136$ & $0.529$ & $0.510$ & $0.075$ \\
                     & $2.20$   & $-15.988$ & $-16.114$ & $3.3067$ & $0.531$ & $0.512$ & $0.076$ \\
                     & $2.00$   & $-15.964$ & $-16.088$ & $3.2524$ & $0.544$ & $0.524$ & $0.077$ \\
                     & $1.88$   & $-15.942$ & $-16.065$ & $3.2027$ & $0.555$ & $0.536$ & $0.078$ \\
                     & $1.80$   & $-15.923$ & $-16.045$ & $3.1600$ & $0.566$ & $0.547$ & $0.079$ \\
\midrule
N\textsuperscript3LO & $\infty$ & $-15.857$ & $-15.903$ & $2.3816$ & $0.765$ & $0.755$ & $0.105$ \\
                     & $2.50$   & $-15.931$ & $-16.026$ & $2.7781$ & $0.642$ & $0.625$ & $0.090$ \\
                     & $2.24$   & $-15.939$ & $-16.043$ & $2.8565$ & $0.623$ & $0.605$ & $0.088$ \\
                     & $2.20$   & $-15.939$ & $-16.044$ & $2.8658$ & $0.621$ & $0.603$ & $0.087$ \\
                     & $2.00$   & $-15.934$ & $-16.043$ & $2.9031$ & $0.614$ & $0.595$ & $0.086$ \\
                     & $1.88$   & $-15.924$ & $-16.034$ & $2.9078$ & $0.615$ & $0.596$ & $0.086$ \\
                     & $1.80$   & $-15.913$ & $-16.023$ & $2.9005$ & $0.618$ & $0.599$ & $0.086$ \\
\midrule
N\textsuperscript4LO & $\infty$ & $-15.870$ & $-15.923$ & $2.5004$ & $0.726$ & $0.715$ & $0.100$ \\
                     & $2.50$   & $-15.929$ & $-16.027$ & $2.8278$ & $0.631$ & $0.614$ & $0.088$ \\
                     & $2.24$   & $-15.934$ & $-16.039$ & $2.8915$ & $0.617$ & $0.598$ & $0.086$ \\
                     & $2.20$   & $-15.934$ & $-16.040$ & $2.8988$ & $0.615$ & $0.597$ & $0.086$ \\
                     & $2.00$   & $-15.926$ & $-16.037$ & $2.9258$ & $0.611$ & $0.592$ & $0.085$ \\
                     & $1.88$   & $-15.915$ & $-16.026$ & $2.9251$ & $0.613$ & $0.594$ & $0.085$ \\
                     & $1.80$   & $-15.903$ & $-16.015$ & $2.9145$ & $0.617$ & $0.598$ & $0.086$ \\
\bottomrule
    \end{tabular}
\end{table}

\begin{table}[H]
    \centering
        \caption{\label{tab:N2LOGO LECs}%
    Same as \cref{tab:ENM LECs} but for the $\Delta\text{N\textsuperscript2LO}_\text{GO}(394)$ interaction.
    The contact term is regularized using $\Lambda=\SI{394}{\MeVc}$ and $n_\text{exp}=4$.
    }
    \begin{tabular}{ccccccc}
\toprule
$\lambda$ & $10^3 \tilde{\mathcal{A}}_L$ & $10^3 \mathcal{A}_L$ & $10^3 \mathcal{A}_S$ & $\tilde{g}$ & $g$ & $\Delta g$ \\
\midrule
$\infty$ & $-15.846$ & $-15.968$ & $3.1225$ & $0.585$ & $0.566$ & $0.080$ \\
$2.50$   & $-15.817$ & $-15.936$ & $3.0524$ & $0.603$ & $0.584$ & $0.082$ \\
$2.24$   & $-15.801$ & $-15.919$ & $3.0155$ & $0.613$ & $0.594$ & $0.083$ \\
$2.20$   & $-15.798$ & $-15.915$ & $3.0088$ & $0.615$ & $0.596$ & $0.083$ \\
$2.00$   & $-15.776$ & $-15.892$ & $2.9610$ & $0.629$ & $0.609$ & $0.084$ \\
$1.88$   & $-15.757$ & $-15.871$ & $2.9212$ & $0.641$ & $0.621$ & $0.086$ \\
$1.80$   & $-15.741$ & $-15.853$ & $2.8882$ & $0.651$ & $0.631$ & $0.087$ \\
\bottomrule
\end{tabular}

\end{table}
\begin{table}[H]
  \centering
    \caption{\label{tab:nme4all}%
    The NME of $\znubb$ decay in $\nuclide[6]{He}$,   $\nuclide[8]{He}$, and $\nuclide[48]{Ca}$, respectively from the calculations with different nuclear chiral potentials. The uncertainties in the $M^{0\nu}_{L+S}$ is propagated from the uncertainty of the $\tilde g$.
  }
  \begin{tabular}{llcccc}
  \toprule
Decay & Interaction & $\lambda ({\rm fm}^{-1})$ &  $M^{0\nu}_L$ & $M^{0\nu}_S/2\tilde g$ & $M^{0\nu}_{L+S}$ \\
  \midrule
 $\nuclide[6]{He}\to\nuclide[6]{Be}$ & EM & 1.8 & 4.017 &0.426 & [4.468, 4.613]\\
   & EMN & 2.0 & 4.101 & 0.510 &[4.640, 4.888] \\
   & LNL & 2.0 & 4.218 & 0.538 & [4.788, 4.971]\\
   & $\Delta$N$^2$LO$_{\rm GO}$ & 2.0 & 4.340 & 0.538 & [4.927, 5.107] \\
   & $\Delta$N$^2$LO$_{\rm GO}$ & $\infty$ &4.221  &0.506 &[4.732, 4.894] \\
  \midrule
 $\nuclide[8]{He}\to\nuclide[8]{Be}$ & EM & 1.8 & 0.129 & 0.158 & [0.297, 0.350] \\
   & EMN & 2.0 & 0.225 & 0.166 &  [0.401, 0.458]\\
   & LNL & 2.0 & 0.165 & 0.176 & [0.352, 0.412] \\
   & $\Delta$N$^2$LO$_{\rm GO}$ & 2.0 & 0.176 & 0.158 & [0.349, 0.402] \\
   & $\Delta$N$^2$LO$_{\rm GO}$ & $\infty$ & 0.168  &  0.157& [0.327, 0.377]\\
  \midrule
 $\nuclide[48]{Ca}\to\nuclide[48]{Ti}$ & EM ($e_{\rm Max}=6$) & 1.8 & 1.03 & 0.357 &  [1.407, 1.529] \\
   & EM ($e_{\rm Max}=8$) & 1.8 & 0.78 & 0.281 &  [1.077, 1.173] \\
   & EM($e_{\rm Max}=10$) & 1.8 & 0.66 & 0.231 & [0.905, 0.983] \\
   & EM(extra.) & 1.8 & 0.61 &  - & [0.836, 0.915] \\
\bottomrule
  \end{tabular}
\end{table}

\end{document}